\documentclass[twocolumn,usenames,dvipsnames]{aastex62}
\pdfoutput=1 
\usepackage{amsmath,amstext}
\usepackage[T1]{fontenc}
\usepackage{apjfonts} 
\usepackage[figure,figure*]{hypcap}
\usepackage{lineno}
\usepackage{verbatim}
\usepackage{fancyhdr}
\usepackage{xcolor}

\shorttitle{High-z Galaxy Clusters in SPT}
\shortauthors{Khullar et al.}

\begin{document}

\title{Spectroscopic Confirmation of Five Galaxy Clusters at $\MakeLowercase{z} > 1.25$ in the 2500 deg$^{2}$ SPT-SZ Survey}

\author[0000-0002-3475-7648]{G. Khullar}
\affiliation{Department of Astronomy and Astrophysics, University of
Chicago, 5640 South Ellis Avenue, Chicago, IL 60637, USA}
\affiliation{Kavli Institute for Cosmological Physics, University of
Chicago, 5640 South Ellis Avenue, Chicago, IL 60637, USA}
\email{Author for correspondence: gkhullar@uchicago.edu}
\author{L.E. Bleem}
\affiliation{Kavli Institute for Cosmological Physics, University of
Chicago, 5640 South Ellis Avenue, Chicago, IL 60637, USA}
\affiliation{Argonne National Laboratory, High-Energy Physics Division,
9700 S. Cass Avenue, Argonne, IL 60439, USA}
\author[0000-0003-1074-4807]{M.B. Bayliss}
\affiliation{Kavli Institute for Astrophysics \& Space Research, Massachusetts Institute of Technology, 77 Massachusetts Ave., Cambridge, MA 02139, USA}
\author[0000-0003-1370-5010]{M.D. Gladders}
\affiliation{Department of Astronomy and Astrophysics, University of
Chicago, 5640 South Ellis Avenue, Chicago, IL 60637, USA}
\affiliation{Kavli Institute for Cosmological Physics, University of
Chicago, 5640 South Ellis Avenue, Chicago, IL 60637, USA} 
\author{B.A. Benson}
\affiliation{Fermi National Accelerator Laboratory, Batavia, IL 60510-
0500, USA}
\affiliation{Department of Astronomy and Astrophysics, University of
Chicago, 5640 South Ellis Avenue, Chicago, IL 60637, USA}
\affiliation{Kavli Institute for Cosmological Physics, University of
Chicago, 5640 South Ellis Avenue, Chicago, IL 60637, USA}
\author{M. McDonald}
\affiliation{Kavli Institute for Astrophysics \& Space Research, Massachusetts Institute of Technology, 77 Massachusetts Ave., Cambridge, MA 02139, USA}
\author{S.W. Allen}
\affiliation{SLAC National Accelerator Laboratory, 2575 Sand Hill Road, Menlo Park, CA 94025, USA}
\affiliation{Kavli Institute for Particle Astrophysics and Cosmology, Stanford University, 452 Lomita Mall, Stanford, CA 94305, USA}
\affiliation{Department of Physics, Stanford University, 382 Via Pueblo Mall, Stanford, CA 94305, USA}
\author{D.E. Applegate}
\affiliation{Kavli Institute for Cosmological Physics, University of
Chicago, 5640 South Ellis Avenue, Chicago, IL 60637, USA}
\author[0000-0002-3993-0745]{M.L.N. Ashby}
\affiliation{Harvard-Smithsonian Center for Astrophysics, 60 Garden Street, MS 66, Cambridge, MA 02138, USA}
\author[0000-0002-4900-805X]{S. Bocquet}
\affiliation{Argonne National Laboratory, High-Energy Physics Division,
9700 S. Cass Avenue, Argonne, IL 60439, USA}
\author{M. Brodwin}
\affiliation{Department of Physics and Astronomy, University of Missouri, 5110 Rockhill Road, Kansas City, MO 64110, USA}
\author[0000-0002-7619-5399]{E. Bulbul}
\affiliation{Harvard-Smithsonian Center for Astrophysics, 60 Garden Street, MS 66, Cambridge, MA 02138, USA}
\author{R.E.A. Canning}
\affiliation{Kavli Institute for Particle Astrophysics and Cosmology, Stanford University, 452 Lomita Mall, Stanford, CA 94305, USA}
\affiliation{Department of Physics, Stanford University, 382 Via Pueblo Mall, Stanford, CA 94305, USA}
\author{R. Capasso}
\affiliation{Faculty of Physics, Ludwig-Maximilians-Universit\"{a}t, Scheinerstr.\ 1, 81679 Munich, Germany}
\affiliation{Excellence Cluster Universe, Boltzmannstr.\ 2, 85748 Garching, Germany}
\author[0000-0002-5819-6566]{I. Chiu}
\affiliation{Academia Sinica Institute of Astronomy and Astrophysics, 11F of AS/NTU Astronomy-Mathematics Building, No.1, Sec. 4, Roosevelt Rd, Taipei 10617, Taiwan}
\author[0000-0001-9000-5013]{T.M. Crawford}
\affiliation{Department of Astronomy and Astrophysics, University of
Chicago, 5640 South Ellis Avenue, Chicago, IL 60637, USA}
\affiliation{Kavli Institute for Cosmological Physics, University of
Chicago, 5640 South Ellis Avenue, Chicago, IL 60637, USA}
\author{T. de Haan}
\affiliation{Department of Physics, University of California, Berkeley, CA 94720, USA}
\author{J.P. Dietrich}
\affiliation{Faculty of Physics, Ludwig-Maximilians-Universit\"{a}t, Scheinerstr.\ 1, 81679 Munich, Germany}
\affiliation{Excellence Cluster Universe, Boltzmannstr.\ 2, 85748 Garching, Germany}
\author{A.H. Gonzalez}
\affiliation{Department of Astronomy, University of Florida, Gainesville, FL 32611, USA}
\author{J. Hlavacek-Larrondo}
\affiliation{Department of Physics, University of Montreal, Montreal, QC H3C 3J7, Canada}
\author{H. Hoekstra}
\affiliation{Leiden Observatory, Leiden University, Niels Bohrweg 2, 2333 CA, Leiden, the Netherlands}
\author{W.L. Holzapfel}
\affiliation{Department of Physics, University of California, Berkeley, CA 94720, USA}
\author{A. von der Linden}
\affiliation{Department of Physics and Astronomy, Stony Brook University, Stony Brook, NY 11794, USA}
\author{A.B. Mantz}
\affiliation{Kavli Institute for Particle Astrophysics and Cosmology, Stanford University, 452 Lomita Mall, Stanford, CA 94305, USA}
\affiliation{Department of Physics, Stanford University, 382 Via Pueblo Mall, Stanford, CA 94305, USA}
\author{S. Patil}
\affiliation{School of Physics, The University of Melbourne, Parkville, VIC 3010, Australia}
\author[0000-0003-2226-9169]{C.L. Reichardt}
\affiliation{School of Physics, The University of Melbourne, Parkville, VIC 3010, Australia}
\author[0000-0002-9288-862X]{A. Saro}
\affiliation{INAF-Observatorio Astronomico di Trieste, via G. B. Tiepolo 11, I-34143 Trieste, Italy}
\author[0000-0002-7559-0864]{K. Sharon}
\affiliation{Department of Astronomy, University of Michigan, 1085 South University Drive, Ann Arbor, MI 48109, USA}
\author{B. Stalder}
\affiliation{LSST, 950 North Cherry Avenue, Tucson, AZ 85719, USA}
\author{S.A. Stanford}
\affiliation{Physics Department, University of California, Davis, CA 95616, USA}
\author{A.A. Stark}
\affiliation{Harvard-Smithsonian Center for Astrophysics, 60 Garden Street, MS 66, Cambridge, MA 02138, USA}
\author{V. Strazzullo}
\affiliation{Faculty of Physics, Ludwig-Maximilians-Universit\"{a}t, Scheinerstr.\ 1, 81679 Munich, Germany}



\begin{abstract}
We present spectroscopic confirmation of five galaxy clusters at $1.25 <$ \textit{z} $< 1.5$, discovered in the 2500 deg$^{2}$ South Pole Telescope Sunyaev-Zel'dovich (SPT-SZ) survey. These clusters, taken from a mass-limited sample with a nearly redshift independent selection function, have multi-wavelength follow-up imaging data from the X-ray to near-IR, and currently form the most homogeneous massive high-redshift cluster sample known. We identify 44 member galaxies, along with 25 field galaxies, among the five clusters, and describe the full set of observations and data products from Magellan/LDSS3 multi-object spectroscopy of these cluster fields. We briefly describe the analysis pipeline, and present ensemble analyses of cluster member galaxies that demonstrate the reliability of the measured redshifts. We report $ \textit{z} =1.259, 1.288, 1.316, 1.401$ and $1.474$ for the five clusters from a combination of absorption-line (Ca II H\&K doublet - $\lambda\lambda$3968,3934\AA) and emission-line ([OII] $\lambda\lambda$3727,3729\AA) spectral features. Moreover, the calculated velocity dispersions yield dynamical cluster masses in good agreement with SZ masses for these clusters. We discuss the velocity and spatial distributions of passive and [OII]-emitting galaxies in these clusters, showing that they are consistent with velocity segregation and biases observed in lower redshift SPT clusters. We identify modest [OII] emission and pronounced CN and H$\delta$ absorption in a stacked spectrum of 28 passive galaxies with Ca II H\&K-derived redshifts. This work increases the number of spectroscopically-confirmed SZ-selected galaxy clusters at $\textit{z}$ > 1.25 from three to eight, further demonstrating the efficacy of SZ selection for the highest redshift massive clusters, and enabling detailed study of these systems.

\end{abstract}



\section{Introduction} \label{introduction}

From overdensities in the initial matter distribution in the Universe, galaxy clusters form and evolve into the massive structures that we observe today. Clusters sample a broad range of galaxy overdensities and mass accretion histories, and studies of these systems provide insight into how stars form and assemble within galaxies, and the evolutionary paths that member galaxies take in cluster environments \citep{1974ApJ...194....1O,1980ApJ...236..351D,1983ApJ...270....7D,1997ApJ...488L..75B,2009ARA&A..47..159B}.

Observations of galaxy clusters at \textit{z} < 1 suggest that galaxies in clusters form stars in an epoch of early and rapid star formation (at \textit{z} > 3), before quickly settling into a mode of passive and stable evolution \citep{1998ApJ...492..461S,2005ApJ...634L.129S,2005ApJ...620L..83H,2006ApJ...644..759M}. Thus, observations of clusters at higher redshifts should sample an epoch where this star formation $-$ or at least its end stages $-$ is observed \textit{in situ}. Recent studies of modest heterogeneous samples of galaxy clusters at 1 < \textit{z} < 2 have shown high star formation and active galactic nuclei (AGN) activity compared with lower redshifts, and a luminosity function that is evolving \citep{2009ApJ...697..436H,2010ApJ...720..284M,2012ApJ...761..141M,2010ApJ...719L.126T,2011NJPh...13l5014F,2012ApJ...756..114S,2012ApJ...756..115Z,2013ApJ...779..138B,2014MNRAS.437..437A,2016ApJ...825...72A}. This is evidence that galaxy clusters are undergoing significant mass assembly in this epoch, inviting further investigation into properties of member galaxies and the intra-cluster medium (ICM) at \textit{z} > 1. 

Although massive clusters are easy to observe in the local universe, the discovery of clusters with similar properties in the high-redshift Universe is still technically challenging. This is due to two main reasons. First, optical and X-ray fluxes -- which are observational tracers of galaxy clusters -- decrease at cosmological distances (due to cosmological dimming). Second, massive galaxy clusters are extremely rare at higher redshifts. Thus, surveys that aim to find distant clusters by directly detecting emission from either the ICM or member galaxies must be both wide and deep. Despite these obstacles, the current status of observations in the \textit{z} > 1 regime is promising, and the science is transforming from the characterization of individual objects to comprehensive analyses of statistically well-defined samples of clusters. A combination of deep X-ray observations \citep{2004AJ....127..230R,2009A&A...508..583R,2005ApJ...623L..85M,2006ApJ...646L..13S,2010ApJ...723L..78C,2018arXiv180307556B} and optical + near-infrared (IR) imaging and spectroscopy \citep{2005ApJ...634L.129S,2012ApJ...753..164S,2014ApJS..213...25S,2006ApJ...651..791B,2011ApJ...732...33B,2006ApJ...639..816E,2006astro.ph..4289W,2008ApJ...684..905E,2009ApJ...698.1934M,2010ApJ...716.1503P,2010ApJ...711.1185D,2011MSAIS..17...66S,2012ApJ...759L..23G,2012ApJ...756..115Z,2015ApJ...812L..40G,2017MNRAS.470.4168B,2017ApJ...844...78P} has improved our understanding of galaxy clusters that have large X-ray, optical, and IR fluxes at 1 < \textit{z} < 2. 

It is also worth noting that these wavelength regimes have their unique advantages. Optical and IR surveys target galaxy overdensities and can probe to low mass thresholds for systems with a breadth of dynamical states and star-formation histories. X-ray observations of clusters provide us with direct measurement of the ICM temperature and electron density, a tracer of cluster mass that is readily captured in cosmological simulations. However, one challenge with optical and IR surveys is whether the selection of galaxy clusters based on galaxies systematically affects the studies of member galaxy properties. To robustly study cluster galaxies absent this concern, an ICM selected sample is appropriate.

Sunyaev-Zel'dovich effect (SZE) cluster surveys from the \textit{Planck} mission \citep{2014AA...571A..20P}, Atacama Cosmology Telescope (ACT, \citealt{2016MNRAS.461..248S,2017arXiv170905600H}), and the South Pole Telescope (SPT,  \citealt{2015ApJS..216...27B}, hereafter B15) offer a new opportunity to study galaxy clusters selected by their ICM signal. Both ACT and SPT provide a nearly-redshift independent, mass-limited sample of clusters, due to their arcminute angular resolution which is well matched to cluster sizes, with a mass-threshold set by the sensitivity of the instruments \citep{2002ARA&A..40..643C}. Of these, only the SPT-SZ cluster catalog yields a significant sample of \textit{z} > 1 clusters. 

The 2500 deg$^{2}$ South Pole Telescope Sunyaev-Zel'dovich (SPT-SZ) survey catalog contains 677 galaxy cluster candidates with a statistical significance > 4.5, with 37 at \textit{z} > 1 (B15) based primarily on photometric red-sequence redshifts. Spectroscopic confirmations along with astrophysical and cosmological analyses of multiple high redshift and massive galaxy clusters from the SPT-SZ survey, many unique to the SPT-SZ sample, have been previously published \citep{2010ApJ...721...90B,2013ApJ...763...93S,2014ApJ...792...45R,2014ApJ...794...12B,2016ApJS..227....3B,2013ApJ...774...23M,2017arXiv170205094M}. This includes spectroscopic confirmation of two particularly distant massive clusters at \textit{z} = 1.322 \citep{2013ApJ...763...93S} and \textit{z} = 1.478 \citep{2014ApJ...794...12B}. This paper provides spectroscopic

\renewcommand{\arraystretch}{1.0}
\startlongtable
\begin{deluxetable*}{cccccc}
\centering
\tablecaption{Galaxy Clusters in the SPT-SZ High-z Cluster Sample\tablenotemark{a} \label{tab:table3}}
\tablehead{
\colhead{Cluster ID} & \colhead{RA} & \colhead{Dec} & \colhead{$\xi\tablenotemark{a}$}& \colhead{Redshift} & \colhead{$M_{500c}$} \\
\colhead{(SPT Cat.)} & \colhead{J2000} & \colhead{J2000} & \colhead{(SZ Significance)} & (Photometric or Previously Published) & \colhead{$10^{14}$ $h_{70}^{-1}$ M$_\odot$}\\
}
\colnumbers
\startdata 
\textbf{SPT-CL J2341-5724} & 355.3568 & $-57.4158$ & 6.87&${1.38\pm0.08}$ & ${3.05\pm0.60}$ \\
\textbf{SPT-CL J0156-5541} & 29.0449 & $-55.6980$ & 6.98&${1.22\pm0.08}$ & ${3.63\pm0.70}$ \\
\textbf{SPT-CL J0640-5113} & 100.0645 & $-51.2204$ & 6.86&${1.25\pm0.08}$ & ${3.55\pm0.70}$ \\
\textbf{SPT-CL J0607-4448} & 91.8984 & $-44.8033$ & 6.44&${1.43\pm0.09}$ & $3.14\pm0.64$\\
\textbf{SPT-CL J0313-5334 }& 48.4809 & $-53.5781$ & 6.09&${1.37\pm0.09}$ & $2.97\pm0.64$\\
\hline
SPT-CL J0205-5829 & 31.4428 & $-58.4852$ & 10.50& $1.322$\tablenotemark{b} & $4.74\pm0.77$\\
SPT-CL J2040-4451 & 310.2483 & $-44.8602$ & 6.72& $1.478$\tablenotemark{c} & $3.33\pm0.66$ \\
SPT-CL J0459-4947 & 74.9269 & $-49.7872$ & 6.29 & ${1.70\pm0.02}$\tablenotemark{d} & $2.67\pm0.55$ \\
\enddata
\tablenotetext{a}{From \citealt{2015ApJS..216...27B}. See Section \ref{sample_sel} for more details.}
\tablenotetext{b}{Spectroscopic follow-up in \citealt{2013ApJ...763...93S}.}
\tablenotetext{c}{Spectroscopic follow-up in \citealt{2014ApJ...794...12B}.}
\tablenotetext{d}{Preliminary result from Mantz et al (in prep).}
\tablecomments{Galaxy clusters in bold are analyzed in this paper.}
\end{deluxetable*}

\noindent confirmation and optical-NIR spectroscopic follow-up of a further five SPT-SZ clusters at 1.25 < \textit{z} < 1.5.

This paper is organized as follows: Section \ref{data} describes the sample selection, optical-NIR imaging, and optical spectroscopy used to derive spectroscopic redshifts for the clusters. In Section \ref{spectra}, we describe the spectral analysis performed on the data from member galaxies of the sample population, while Section \ref{data_products} describes the resulting spectroscopic redshifts and confirmation of member galaxies. In Section \ref{discussion}, we consider several analyses of these data - cluster velocity dispersions, a stacked velocity-radius diagram, and a stacked spectral analysis, all of which demonstrate -- despite the challenge presented by spectroscopy of individual member galaxies in these distant systems -- that the spectroscopic results are as expected. We summarize our results in Section \ref{summary}.

Magnitudes in this work have been calibrated with respect to Vega. The fiducial cosmology model used for all distance measurements as well as other cosmological values assumes a standard flat cold dark matter universe with a cosmological constant $(\Lambda$CDM), H$_0$ = 70 km s$^{-1}$ Mpc$^{-1}$, and matter density $\Omega_M$ = 0.30. All Sunyaev-Zel'dovich (SZ) significance-based masses from B15 are reported in terms of  M$_{500c,SZ}$ i.e. the SZ mass within R$_{500c}$, defined as the radius within which the mean density $\rho$ is 500 times the critical density $\rho_{c}$ of the universe. 


\section{Observations and Data} \label{data}

\subsection{Cluster Sample Selection and Imaging Follow-up} \label{sample_sel}

The 2500 deg$^{2}$ SPT-SZ survey \citep{2011PASP..123..568C}, completed in 2011, discovered 37 galaxy clusters with high significance at \textit{z} > 1, via the SZE. These clusters were detected via the SPT-SZ campaign that observed the CMB at frequencies 95, 150, and 220 GHz. The full cluster catalog, B15, is $\sim$ 100\% complete at \textit{z} $>$ 0.25 for a mass threshold of M$_{500c}$ $\geq$ 7 $\times 10^{14}$ M$_\odot$ $h^{-1}_{70}$. Survey strategy and analysis details can be found in previous work by the SPT collaboration \citep{2009ApJ...701...32S,2010ApJ...722.1180V,2011ApJ...738..139W,2013ApJ...763..127R}.

\begin{figure*}
\epsscale{1.00}
\plotone{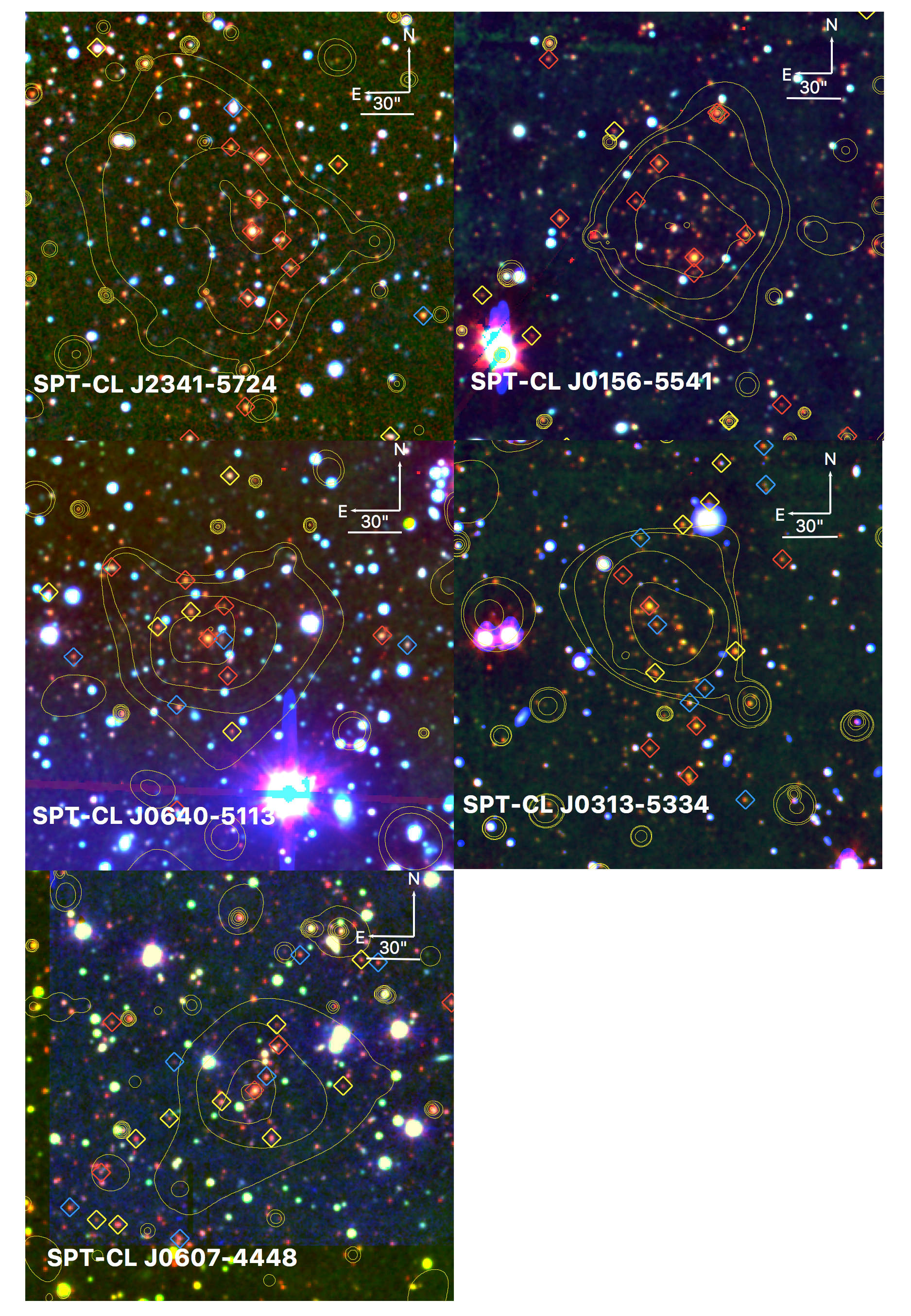}
\caption{RGB $4\arcmin \times 4\arcmin$ images for the sample clusters. Data are {\sl Spitzer} IRAC 3.6\micron\ (red channel), Magellan/FOURSTAR J,H or Ks (green channel), and Magellan/PISCO z (blue channel), except for SPT-CL J0607-4448, for which the blue channel is ESO/NTT \textit{z} band data. Images are centered on their SZ centers. Diamonds indicate all objects targeted for spectroscopic observations. Red diamonds indicate spectroscopically confirmed cluster members (see Sections \ref{spectra} and \ref{data_products}). Yellow diamonds indicate objects for which no redshifts could be measured, while blue diamonds indicate confirmed field galaxies. Contours are drawn from smoothed {\sl Chandra} X-ray data for these clusters \citep{2017arXiv170205094M}, spaced equally in log$_{10}$(flux) from the lowest discernible value that isolate the cluster, up to just beyond the peak of the diffuse emission from the cluster. 
\label{fig:rgb1}}
\end{figure*}

For optical and near-IR (NIR) photometric follow-up of this cluster sample, several programs were initiated (see \citealt{2012ApJ...761...22S} and B15 for details on observational strategies). Optical photometry in the $\it{griz}$ bands was obtained for the sample clusters using either the CTIO (4m) facility, ESO/New Technology Telescope (NTT, 3.58m) or the Magellan/Baade Telescope (6.5m) in Chile, to depths that can detect galaxies at 0.4L$\ast$ at \textit{z} = 0.75 at 5$\sigma$ in red bands. This was followed up by Spitzer/IRAC observations in the 3.6$\mu$m and 4.5$\mu$m bands for NIR photometry, which is crucial for observations of higher redshift clusters for member galaxy candidate selection. The final NIR images detect \textit{z} = 1.5 0.4L$^{\ast}$ galaxies at a 10$\sigma$ significance.

The IR and optical photometry is complemented by observations in bands $J, H, H-$long and $K_s$ bands (modified versions of the standard $H$ and $K$ filters, respectively), using the wide-area near-IR instrument FOURSTAR on the Magellan/Baade telescope \citep{2013PASP..125..306F}. Data were acquired between January 2014 $-$ January 2016; the data used here are a small subset of the overall dataset, with details of the reduction and analysis to be provided in a future paper \citetext{Bayliss et al., in prep}. Deep optical photometric follow-up was also acquired for four of the five clusters discussed in this paper using the simultaneous \textit{griz} imager Magellan/PISCO \citep{2014SPIE.9147E..3YS} on 2 November 2016. Deep \textit{HST}/WFC3 photometry in the F814W and F140W filters (PI: Strazzullo, HST Cycle 23 program) is available for the fifth cluster, SPT-CL J0607-4448. RGB images for the sample clusters are shown in Figure \ref{fig:rgb1}.

The cluster sub-sample analyzed in detail here comprises five of the eight most massive \textit{z} > 1.2 clusters from the SPT-SZ sample, all of which have deep \textit{Chandra} X-ray imaging \citep{2017arXiv170205094M}. The remaining three are SPT-CL J2040-5541 (spectroscopically confirmed at z = 1.478 in \citealt{2014ApJ...794...12B}), SPT-CL J0205-5829 (spectroscopically  confirmed at \textit{z} = 1.322 in \citealt{2013ApJ...763...93S}), and SPT-CL J0459-4947, for which current data provides an X-ray spectroscopic redshift of $1.70\pm0.02$ \citetext{Mantz et al., in prep, with a past published redshift of 1.85 in \citealt{2017arXiv170205094M}}. This total sample is referred to as the "SPT High-z Cluster" sample.

In Section \ref{spec_oir}, we describe the spectroscopic optical observations of this five cluster sub-sample. The cataloged properties of these five clusters, and the further three systems that complete the set of the most massive high-redshift clusters in the SPT-SZ sample, are reproduced from B15 in Table \ref{tab:table3}. The photometric data used for constructing RGB images in this work are summarized in Table \ref{tab:table_photometry}.

\subsection{Spectroscopy: Optical and Near-IR}\label{spec_oir}

\subsubsection{Spectral Observations}

The primary motivation of optical and NIR follow-up of this cluster sample is securing spectroscopic redshifts of clusters and their member galaxies. Optical spectroscopy of these five clusters was carried out between August 2014 and January 2015 on the 6.5m Magellan/Clay Telescope using the 600 lines/mm VPH-Red grism on the Low Dispersion Survey Spectrograph\footnote{http://www.lco.cl/Members/gblanc/ldss-3/ldss-3-user-manual-tmp} $-$ 3C (LDSS3C) in Normal mode (as opposed to nod-and-shuffle). These data represent some of the earliest spectroscopy acquired with the new  LDSS3C system, and include both unfiltered spectra, and spectra acquired using the OG590 order separating filter - the latter being used to remove second order contamination in cluster spectra where imaging showed higher blue-end flux.

The slits for target galaxy spectra were typically cut 6" long (along the spatial axis) on the mask and 1" wide (along the dispersion axis); LDSS3C has a scale of 0.188"/pixel. In most instances, the target galaxy was positioned at the slit center, with some misalignment on the spatial axis tolerated in order to optimize slit packing. Square boxes, typically 6 per mask, were used to target nearby stars for mask alignment on the sky. Spectra of individual galaxies typically cover the wavelength range $7500-10000${\AA}, with a typical exposure time of 7200s and an observation airmass of $\sim1.2-1.5$. The typical seeing during the observations was $\sim$ 1".

\renewcommand{\arraystretch}{1.0}
\startlongtable
\begin{deluxetable}{c c}
\setlength{\tabcolsep}{10pt}
\centering
\tablecaption{Photometric Data in this study \label{tab:table_photometry}}
\tablehead{
\colhead{Cluster Name}&\colhead{Imaging (RGB)}\\
\colhead{} & \colhead{(Telescope and Instrument)}\\
}
\colnumbers
\startdata 
SPT-CL J2341-5724 & Spitzer/IRAC 3.6$\mu$m\\
& Magellan/FOURSTAR J\\
& Magellan/PISCO z\\
SPT-CL J0156-5541 & Spitzer/IRAC 3.6$\mu$m\\
& Magellan/FOURSTAR H\\
& Magellan/PISCO z\\
SPT-CL J0640-5113 & Spitzer/IRAC 3.6$\mu$m\\
& Magellan/FOURSTAR J\\
& Magellan/PISCO z\\
SPT-CL J0607-4448 & Spitzer/IRAC 3.6$\mu$m\\
& Magellan/FOURSTAR J\\
& ESO/NTT z\\
SPT-CL J0313-5334 & Spitzer/IRAC 3.6$\mu$m\\
& Magellan/FOURSTAR Ks\\
& Magellan/PISCO z\\
\enddata
\tablecomments{See Section \ref{sample_sel} for more details on imaging follow-up of sample clusters.}
\end{deluxetable}

\begin{figure*}[t!]
\epsscale{1.05}
\plotone{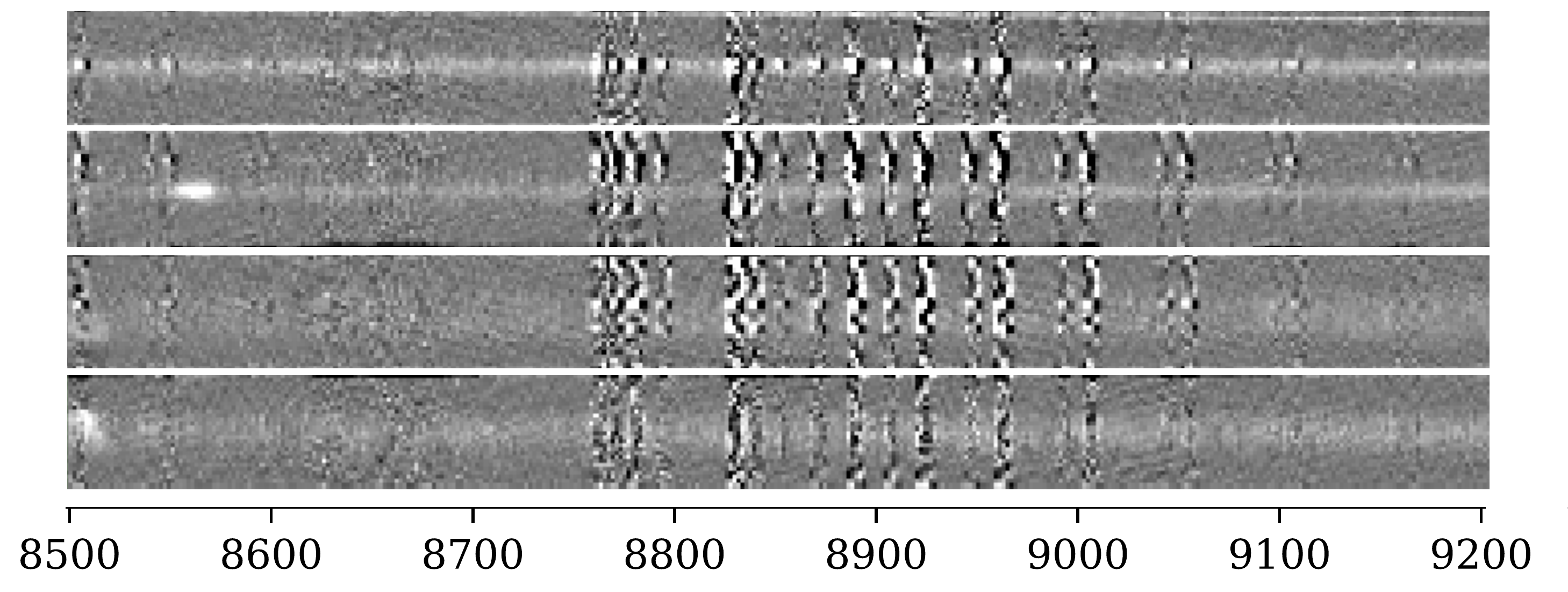}
\caption{Example of 2D spectra used in the analysis described in Section \ref{spectra}, for four candidate member galaxies of the cluster SPT-CL J0156-5541. The vertical direction is the spatial dimension, and the horizontal direction is the dispersion axis, which runs from $8500-9200$\AA\ from left to right. Emission features ([OII]) can be seen in the 8500-8600\AA\ region.
\label{fig:2dspec}}
\end{figure*}

\subsubsection{Designing Spectroscopic Masks}
Likely  bright red-sequence cluster members, selected by apparent color, morphology, and proximity to the SZ center and corresponding galaxy overdensity, were used to choose an appropriate mask orientation on the sky, and were targeted first for slit placement. Further slits were placed on fainter or bluer galaxies, as allowed by the mask geometry, until no more slits could be placed. Approximately 20-25 targets were chosen for each spectroscopic mask. 
As noted above, some spatial misalignment ($\sim 1/10$th of an arcsecond) was allowed to optimize slit packing. This process typically results in an elongated (rather than circular) layout of targeted galaxies, as can be seen in Figure \ref{fig:rgb1}. Care was taken to place a slit on any apparent brightest cluster galaxy (BCG). The visual mask design was then used to generate an input catalog from the slit positions, which was then input to the standard LDSS3C mask design software to produce the final mask design. 

\section{Spectral Analysis} \label{spectra}
\subsection{Spectra Reduction}

The spectra were processed using The Carnegie Observatories System for Multi-Object Spectroscopy (COSMOS)\footnote{http://code.obs.carnegiescience.edu/cosmos} reduction package, which is specifically designed to reduce raw spectra acquired using the Magellan Telescopes.

We describe the data reduction briefly below. All images were de-biased using bias frames aquired each afternoon. LDSS3C has a modest pattern noise of 1-2 electrons amplitude and care was taken to acquire sufficient bias frames to average across this noise source. We used a HeNeAr comparison arc line for wavelength calibration. The analysis is focused on the range where the VPH-red grism is most sensitive and over which we expect useful spectral signal from these very red member galaxies : 7500-10000\AA. A flat field image acquired temporally adjacent to each science frame was used to define the spectral trace for each slit. This flat image was also used to flat-field the slit response. Sky subtraction was performed by fitting a one-dimensional third-order spline along the dispersion axis, following the techniques outlined in \cite{2003PASP..115..688K}. Different exposures of the sky-subtracted science spectra were stacked and 2D cosmic ray cleaning was performed by outlier rejection. COSMOS also generates a noise image, that, at these red wavelengths, is dominated by photon noise in bright sky lines.

Figure \ref{fig:2dspec} shows examples of 2-D sky-subtracted spectra from 4 potential member galaxies of the galaxy cluster SPT-CL 0156-5541. The y-axis depicts the spatial width of individual slits, against the horizontal dispersion axis (or wavelength), over the wavelength range 8500-9200\AA. Emission features ([OII] $\lambda$3727,2729\AA\ doublet) and a strong spectral continuum can be clearly seen in some cases. It is crucial to note that there are sections of the spectra in the wavelength range of interest that are dominated by poor sky subtraction with significant systematics. A thorough exploration of the tunable parameters available in the COSMOS reduction package did little to ameliorate this $-$ the poor sky subtraction is not due to an insufficient description of the slit geometry. In principle, fringing could contribute to this effect, but the LDSS3C detector is a thick fully depleted CCD, the same as the chips used in the Dark Energy Camera \citep{2015AJ....150..150F}, and is not expected to show fringing at this level. Discussions with several architects of the COSMOS reduction code have led to the conclusion that these artifacts are predominantly the result of slit roughness, that does not fully flat-field away because the effect produces a transmission variation along the slit spatial axis, as well as a high-frequency variation in the wavelength solution. The latter is not addressed by flat-fielding, and removing such artifacts requires a reworking of the COSMOS sky-subtraction engine. The quality of 2D spectra is a major factor in selecting analysis strategies for 1D spectra, which are described in the following sections. 

\renewcommand{\arraystretch}{1.5}
\startlongtable
\begin{deluxetable*}{c c c c c c }
\setlength{\tabcolsep}{9pt}
\tabletypesize{\small}
\centering
\tablecaption{Spectroscopic Redshifts of Member Galaxies\tablenotemark{a} \label{tab:table_all}}
\tablehead{
\colhead{Cluster Name} & \colhead{Galaxy RA}& \colhead{Galaxy Dec}& \colhead{$z$} & \colhead{$\delta z$} & \colhead{Principal Spectral Feature} \\ 
\colhead{} & \colhead{(J2000)}&\colhead{(J2000)} & \colhead{(RVSAO+[OII]+customcode)} & &
}
\colnumbers
\startdata 
SPT-CL J2341-5724& $23:41:24.792$ &$-57:25:01.25$&1.2570 & 0.0010 & Ca II H\&K \\
SPT-CL J2341-5724& $23:41:24.077$ &$-57:24:19.71$&1.2610 & 0.0020 & Ca II H\&K \\
SPT-CL J2341-5724& $23:41:29.282$ &$-57:26:56.12$&1.2550 & 0.0010 & Ca II H\&K \\
SPT-CL J2341-5724& $23:41:24.277$ &$-57:24:43.50$&1.2582 & 0.0020 & Ca II H\&K \\
SPT-CL J2341-5724& $23:41:25.135$ &$-57:25:38.42$&1.2501 & 0.0008 & Ca II H\&K \\
SPT-CL J2341-5724& $23:41:25.396$ &$-57:26:38.72$&1.2510 & 0.0030 & Ca II H\&K  \\
SPT-CL J2341-5724& $23:41:23.082$ &$-57:25:50.93$&1.2581 & 0.0004 & Ca II H\&K \\
SPT-CL J2341-5724& $23:41:22.732$ &$-57:25:06.47$&1.2638 & 0.0016 & Ca II H\&K \\
SPT-CL J2341-5724& $23:41:26.185$ &$-57:24:14.38$&1.2687 & 0.0008 & Ca II H\&K \\
SPT-CL J2341-5724& $23:41:22.169$ &$-57:25:21.68$&1.2701 & 0.0006 & Ca II H\&K \\
\hline
SPT-CL J0156-5541& $01:56:09.109$ & $-55:42:10.51$ &1.2935 & 0.0015 & Ca II H\&K\\
SPT-CL J0156-5541& $01:56:03.382$ &$-55:43:32.36$&1.2877 & 0.0050 & [OII] \\
SPT-CL J0156-5541& $01:56:18.665$ &$-55:40:20.74$&1.2825 & 0.0011 & Ca II H\&K / [OII]\\
SPT-CL J0156-5541& $01:56:11.439$ &$-55:41:18.49$&1.2925 & 0.0030 & Ca II H\&K \\
SPT-CL J0156-5541& $01:56:05.725$ &$-55:41:57.81$&1.2802 & 0.0009 & Ca II H\&K \\
SPT-CL J0156-5541& $01:56:09.134$ &$-55:42:19.08$&1.2970 & 0.0030 & Ca II H\&K / [OII] \\
SPT-CL J0156-5541& $01:56:12.938$ &$-55:41:39.55$&1.2925 & 0.0020 & Ca II H\&K \\
SPT-CL J0156-5541& $01:56:17.928$ &$-55:41:49.02$&1.2980 & 0.0020 & Ca II H\&K \\
SPT-CL J0156-5541& $01:55:55.826$ &$-55:43:10.49$&1.2830 & 0.0001 & Ca II H\&K / [OII] \\
SPT-CL J0156-5541& $01:56:07.627$ &$-55:40:50.23$& 1.2772 & 0.0005 & Ca II H\&K / [OII] \\
SPT-CL J0156-5541& $01:56:07.438$ &$-55:40:51.17$& 1.2810\tablenotemark{b}  & 0.0010 & [OII] \\ 
SPT-CL J0156-5541& $01:55:59.089$ &$-55:43:49.65$ &1.2900 & 0.0010 & Ca II H\&K \\
SPT-CL J0156-5541& $01:55:56.724$ &$-55:39:27.54$&1.2832 & 0.0007 & Ca II H\&K / [OII] \\
SPT-CL J0156-5541& $01:56:11.064$ &$-55:38:31.06$&1.2841 & 0.0020 & Ca II H\&K\\
SPT-CL J0156-5541& $01:56:05.607$ &$-55:38:42.17$&1.2970 & 0.0005 & Ca II H\&K / [OII]\\
\hline
SPT-CL J0640-5113& $06:40:17.377$ &$-51:13:04.04$&1.3180 & 0.0014 & Ca II H\&K \\
SPT-CL J0640-5113& $06:40:18.690$ &$-51:12:31.81$&1.3120 & 0.0002 & Ca II H\&K \\
SPT-CL J0640-5113& $06:40:23.045$ &$-51:12:24.57$&1.3264 & 0.0010 & Ca II H\&K \\
SPT-CL J0640-5113& $06:40:16.204$ &$-51:13:24.86$&1.3031 & 0.0002 & Ca II H\&K / [OII] \\
SPT-CL J0640-5113& $06:40:07.080$ &$-51:13:02.32$&1.3209 & 0.0020 & Ca II H\&K \\
SPT-CL J0640-5113& $06:40:16.400$ &$-51:12:46.13$&1.3210 & 0.0020 & Ca II H\&K \\
SPT-CL J0640-5113& $06:40:19.194$ &$-51:14:39.36$&1.3079 & 0.0002 & Ca II H\&K / [OII] \\
\hline
SPT-CL J0607-4448& $06:07:34.218$ &$-44:48:07.30$&1.4087& 0.0010 & Ca II H\&K \\
SPT-CL J0607-4448& $06:07:32.462$ &$-44:46:59.70$&1.4077& 0.0008 & Ca II H\&K \\
SPT-CL J0607-4448& $06:07:38.712$ &$-44:49:36.72$&1.3973& 0.0006 & Ca II H\&K \\
SPT-CL J0607-4448& $06:07:44.442$ &$-44:49:19.70$&1.3948& 0.0012 & Ca II H\&K \\
SPT-CL J0607-4448& $06:07:34.824$ &$-44:48:14.95$&1.3993\tablenotemark{c}& 0.0013 & [OII] \\
\hline
SPT-CL J0313-5334& $03:13:58.536$ &$-53:32:31.50$&1.4695 & 0.0001 & [OII] \\
SPT-CL J0313-5334& $03:13:48.216$ &$-53:33:48.60$&1.4740 & 0.0010 & Ca II H\&K / [OII] \\
SPT-CL J0313-5334& $03:13:58.105$ &$-53:33:57.30$&1.4881& 0.0004&[OII]\\
SPT-CL J0313-5334& $03:13:56.472$ &$-53:34:14.61$&1.4772& 0.0010&Ca II H\&K\\
SPT-CL J0313-5334& $03:13:53.569$ &$-53:35:21.12$&1.4730& 0.0010&Ca II H\&K\\
SPT-CL J0313-5334& $03:13:56.448$ &$-53:35:33.50$&1.4716& 0.0006&Ca II H\&K / [OII]\\
SPT-CL J0313-5334& $03:13:54.049$ &$-53:35:49.01$ &1.4770& 0.0008&Ca II H\&K / [OII]\\
\hline
\enddata
\tablenotetext{a}{From a combination of RVSAO cross-correlation and fit to [OII] emission features.}
\tablenotetext{b}{Second trace of galaxy that fell serendipitously into the slit. Possible member galaxy.}
\tablenotetext{c}{BCG of SPT-CL J0607-4448, confirmed after revisiting the spectrum (See Section \ref{curiouscase} for details).}
\tablecomments{Spectroscopic redshifts of member galaxies of sample clusters, in increasing order of the cluster redshifts. Also mentioned are the spectral features used to determine each galaxy's redshift. See Section \ref{spectra} and \ref{data_products} for more details.}
\end{deluxetable*}

\begin{figure*}[t!]
\plotone{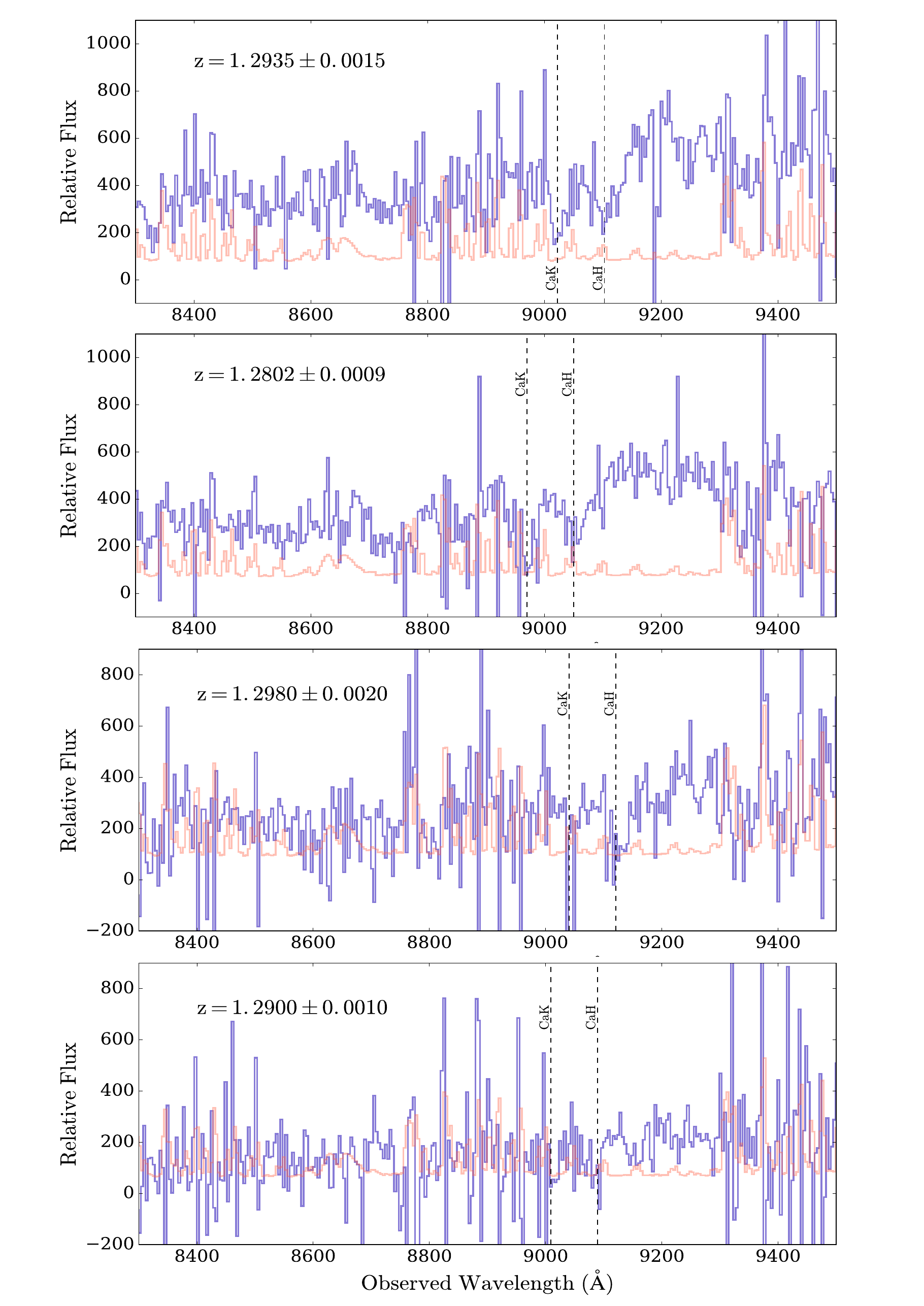}
\caption{Extracted 1-D spectra (purple) for 4 member galaxies of the cluster SPT-CL J0156-5541 (see Table \ref{tab:table_all}), along with 1$\sigma$ uncertainties (orange). Ca II H\&K absorption features are indicated (black dashed lines), corresponding to robust redshift fits from RVSAO cross-correlation. Redshift values reported in this figure are final redshifts from the combined RVSAO and line identification analyses.
\label{fig:general3}}
\end{figure*}

\subsection{One-dimensional Spectra}
Two-dimensional spectra are condensed into one dimensional spectra for analysis using the IRAF/NOAO package \textit{apall} that fits polynomial functions to the spectral continuum (along the dispersion axis) in individual 2D spectra. Along the spatial axis, the process involved clipping slit edges for defects, and fitting a boxcar model to the counts distribution. At any wavelength, the RMS of the sky subtracted residuals defines the uncertainties.

\subsection{Extracting Redshifts and Spectral Features}

Due to sub-optimal sky subtraction in the LDSS3 pipeline, the resulting 1D wavelength-calibrated (albeit not flux-calibrated) spectra are dominated by systematic noise at some wavelengths, and not suitable for sophisticated spectral analysis techniques that can be employed to analyze galaxy spectra (e.g. principal-component analysis). Several strong spectral features are apparent in some spectra, and we base much of the analysis that follows on the detection of these features, namely the [OII] $\lambda\lambda\ 3727,3729${\AA} doublet emission lines, and Ca II H\&K $\lambda\lambda\ 3968,3934${\AA} absorption lines. In the case of some passive galaxies, a modest H$\delta\ 4102${\AA} line may also be observed corresponding to the Ca II H\&K-based redshifts, but not extracted independently. These spectral features were first identified visually in both 2D and 1D spectra, and analysed using two separate methods that are described below. It is important to note that spectral features and redshifts can robustly be identified without flux calibration of spectra (see further discussion in Section \ref{discussion}).

\subsubsection{Redshifts from Cross-Correlation: RVSAO}

We use the Harvard Smithsonian Astrophysical Observatory's Radial Velocity (RVSAO) IRAF package \citep{1998PASP..110..934K} to implement a cross-correlation analysis between our wavelength-calibrated 1D spectrum and a galaxy template spectrum. To this end, we employed a standard template, $\it{fabtemp97}$, that contains absorption features commonly seen in spectra of cluster member galaxies. For the low S/N data at our disposal, we use the Ca II H\&K absorption lines at rest-frame wavelengths of  $\lambda3968,3934${\AA} which fall in the observer-frame wavelength range of 8800-9400{\AA} for the redshifts of our sample clusters. Challenges in obtaining the redshift solutions for our dataset via this method are further discussed in Section \ref{data_products}.

\begin{figure*}[t!]
\epsscale{1.13}
\plotone{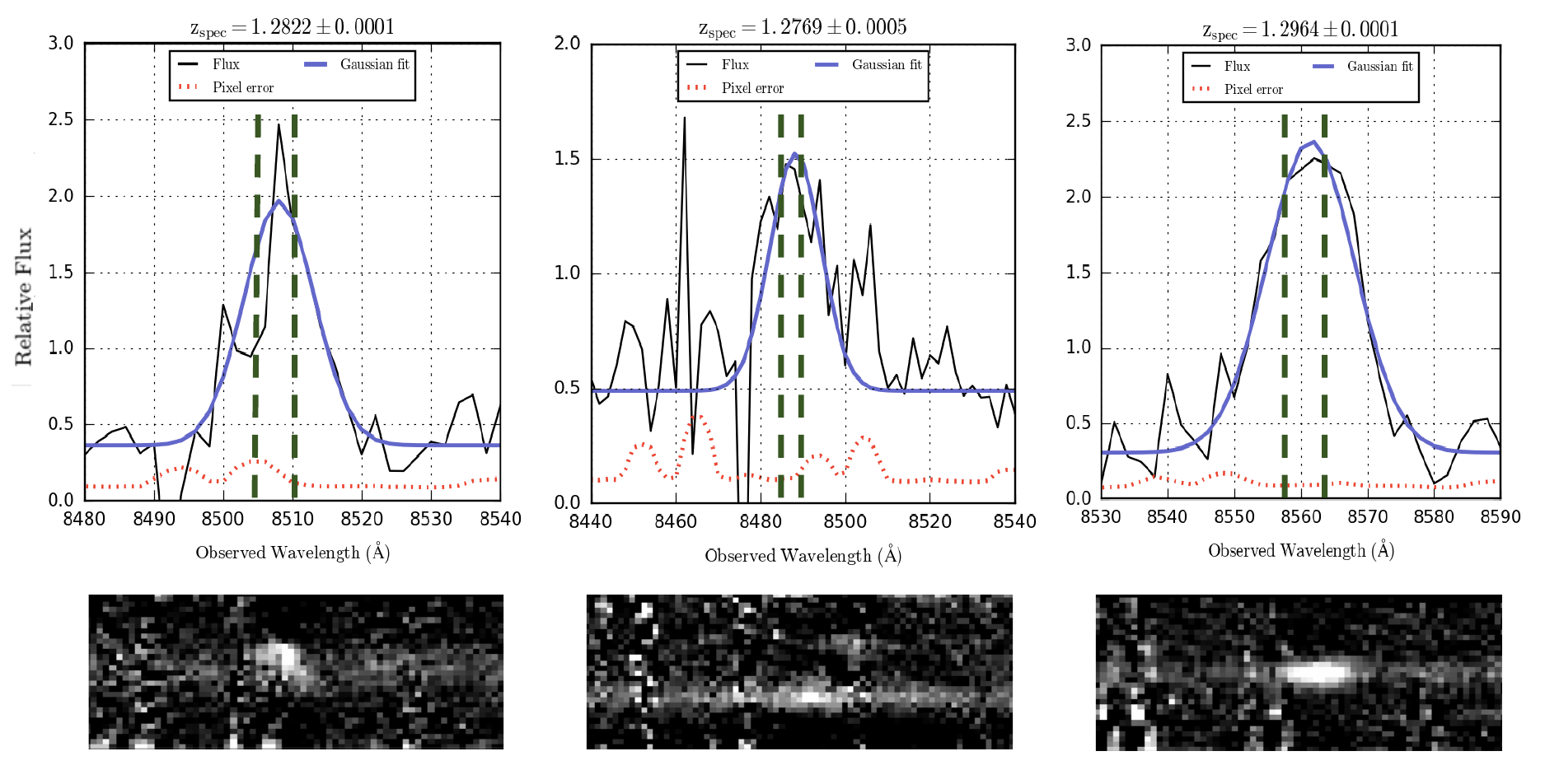}
\caption{{\sl Upper panels}: Examples of individual 1D spectra (black solid line) for individual member galaxies of SPT-CL J0156-5541 (see Table \ref{tab:table_all}) that show strong [OII] emission lines. Uncertainties are indicated with red dotted lines. The main emission feature reflects the emission from the [OII] 3727,3729\AA\ doublet, which is blended at our resolution. The mean redshift (corresponding to a rest-frame wavelength of 3728.1\AA) is used to constrain spectroscopic redshifts for these galaxies. The dashed vertical lines indicate the observed wavelengths of the two lines in the [OII] doublet. The purple curves are Gaussian fits to the emission features. {\sl Lower panels}:  2D spectra corresponding to the galaxies above. The lower-central panel contains two [OII]-emitting cluster members serendipitously observed with the same slit.}
\end{figure*}

\subsubsection{Redshifts from Line Identification}

In order to estimate redshifts as an independent probe of low-S/N, sparsely featured spectra, and to substantiate our RVSAO redshift measurements in moderate- and high- S/N spectra, we use the detections of [OII] $\lambda\lambda$ 3727,3729{\AA} doublet emission features, which fall in the observer-frame wavelength range of 8300-8700{\AA} for the redshifts of our sample clusters. Since the dispersion of our 1D spectra is 2{\AA}/pixel, we do not expect to resolve individual lines in this doublet feature. However, the width of the features we identify as [OII] emission is consistent with a redshifted blended [OII] doublet line profile. In our analysis, we also consider the uncertainty -- albeit typically subdominant -- in the median wavelength of the blended [OII] doublet feature, due to the range in the [OII] doublet line ratio from varying physical conditions.
In most cases where a single emission feature is used to characterize the galaxy redshift, we are able to visually confirm a 4000\AA\ break at the observed wavelength corresponding to the redshift candidate. In one case (a galaxy observed within the field-of-view of SPT-CL J0156-5541), this clear a diagnosis was not possible i.e. the emission feature could potentially correspond to an [OIII], H$\beta$ or [OII] emission peak. [OIII] was disfavored because it is generally accompanied by a blueward H$\beta$ peak, that was not observed. H$\beta$ was ruled out because a nominal redward [OIII] peak was not seen, and the existence of an [OII] peak corresponding to the cluster redshift (confirmed by 14 other cluster members in the field-of-view) increased our confidence in this feature being attributed to [OII].

We also independently analyzed our sample spectra using a custom IDL code (from here on, \textit{customcode}) designed to help identify multiple low S/N spectral features. We visually examined each spectrum, with typically multiple redshift solutions considered to fit apparent spectral features present in the data. Final redshifts were derived from the median of the individual line fits, with the variance providing an estimate of redshift uncertainties.

We discuss the methodology of calculating uncertainties (provided in Table \ref{tab:table_all}) in Section \ref{unc}.

\section{Data Products and Results} \label{data_products}
\subsection{Spectroscopic redshifts of member galaxies}

Table \ref{tab:table_all} contains galaxy coordinates and spectroscopic redshifts for all galaxies being considered as member galaxies for our 5 sample clusters. Also mentioned are the spectral features that were used to characterize the redshift. The total number of target galaxies upon which slits were placed is 109, excluding objects that serendipitously fell onto the slit. Of these, we consider 39 redshifts to be of high confidence. In addition, we include 4 galaxies with lower confidence redshifts, that correspond to measurements with higher uncertainties than are usual for spectroscopic redshifts for galaxies ($\Delta$\textit{z} > 0.002). We also include the moderately robust redshift for the brightest cluster galaxy (BCG) in SPT-CL J0607-4448 (see Section \ref{curiouscase} for a detailed discussion). Inclusion of these 5 galaxies does not affect the scientific results of this analysis.

Figure \ref{fig:general3} shows examples of one-dimensional spectra of 4 member galaxies of SPT-CL J0156-5541 at \textit{z} = 1.2935, 1.2802, 1.2980 and 1.2900, along with 1$\sigma$ pixel errors as a function of wavelength. The absorption features corresponding to Ca II H\&K can be observed, indicating the presence of a dominant older stellar population.

\renewcommand{\arraystretch}{1.0}
\startlongtable
\begin{deluxetable*}{c c c c c c }
\centering
\tablecaption{Spectroscopic Redshifts of Field Galaxies in the Dataset\tablenotemark{a} \label{tab:table_nonmembers}}
\tablehead{
\colhead{Spectroscopic Mask ID} & \colhead{Galaxy RA}& \colhead{Galaxy Dec}& \colhead{$z$} & \colhead{$\delta z$} & \colhead{Principal Spectral Feature} \\
\colhead{} & \colhead{(J2000)}&\colhead{(J2000)} & & & 
}
\colnumbers
\startdata 
SPT-CL J2341-5724& $23:41:13.112$ & $-57:25:49.46$ &1.3410 & 0.0100 & Ca II H\&K\\
SPT-CL J2341-5724& $23:41:25.993$ & $-57:23:52.63$ &1.3272\tablenotemark{b} & 0.0005 & Ca II H\&K\\
SPT-CL J2341-5724& $23:41:23.041$ & $-57:27:34.72$ &1.1436 & 0.0017 & Ca II H\&K / [NeIII] \\
SPT-CL J2341-5724& $23:41:23.686$ & $-57:22:32.03$ &0.8031 & 0.0006 & Ca II H\&K  \\
SPT-CL J0640-5113& $06:40:25.283$ & $-51:13:14.46$ &0.6404 & 0.0001 & [OIII]\\
SPT-CL J0640-5113& $06:40:05.611$ & $-51:13:07.66$ &0.8189 & 0.0002 & H$\beta$ / [OIII]\\
SPT-CL J0640-5113& $06:40:16.445$ & $-51:13:04.79$ &2.4840 & 0.0010 & FeII / MgII\\
SPT-CL J0640-5113& $06:40:19.209$ & $-51:13:41.13$ &1.3590 & 0.0010 & Ca II H\&K\\
SPT-CL J0607-4448& $06:07:33.586$ & $-44:47:49.66$ &1.4933&0.0005&Ca II H\&K\\
SPT-CL J0607-4448& $06:07:38.992$ & $-44:47:59.12$ &1.7181&0.0004&[OII]\\
SPT-CL J0607-4448& $06:07:24.579$ & $-44:47:26.57$ &1.4716&0.0011&Ca II H\&K\\
SPT-CL J0607-4448& $06:07:28.380$ & $-44:47:03.95$ &1.3078&0.0013&Ca II H\&K\\
SPT-CL J0607-4448& $06:07:42.238$ & $-44:47:37.27$ &1.4787&0.0004&[OII]\\
SPT-CL J0607-4448& $06:07:42.844$ & $-44:48:59.94$ &1.4965&0.0004&[OII] / H$\delta$\\
SPT-CL J0313-5334& $03:13:49.369$ & $-53:32:45.96$ &1.0926 & 0.0008 & Ca II H\&K / G\\
SPT-CL J0313-5334& $03:13:49.369$ & $-53:32:45.96$&0.86851\tablenotemark{b} & 0.0001 & [OIII] / H$\beta$\\
SPT-CL J0313-5334& $03:13:49.248$ & $-53:33:07.39$ &1.0680&0.0001 & Hg / H$\beta$ / [OIII]\\
SPT-CL J0313-5334& $03:13:49.248$ & $-53:33:07.39$&1.3586\tablenotemark{b} & 0.0003 & [OII]\\
SPT-CL J0313-5334& $03:13:57.024$ & $-53:33:36.79$ &1.3029&0.0008& [OII]\\
SPT-CL J0313-5334& $03:13:56.472$ & $-53:34:14.611$&1.2320&0.0010& [OII]\\
SPT-CL J0313-5334& $03:13:55.993$ & $-53:34:24.96$ &1.2313&0.0003& [OII] / H$\delta$\\
SPT-CL J0313-5334& $03:13:53.040$ & $-53:35:00.24$ &1.2150&0.0010& [OII]\\
SPT-CL J0313-5334& $03:13:53.982$ & $-53:35:08.11$ &1.2620&0.0004& [OII]\\
SPT-CL J0313-5334& $03:13:50.521$ & $-53:36:02.16$ &6.1480&0.0010& Ly$\alpha$\\
SPT-CL J0313-5334& $03:13:56.017$ & $-53:36:59.91$ &1.1591&0.0006& Ca II H\&K\\
\enddata
\tablenotetext{a}{From a combination of RVSAO cross-correlation and line identification.}
\tablenotetext{b}{A second trace of a galaxy that fell serendipitously onto the slit.}
\end{deluxetable*}

Some of the spectra correspond to non-member galaxies (foreground or background) as well as stars. We describe non-member galaxy spectra in Table \ref{tab:table_nonmembers}. It can be seen that some of these field galaxies have strong forbidden line features (e.g. NeIII) that are associated with AGN activity. This list of galaxies also includes a \textit{z} = 2.48 background galaxy with potential FeII/MgII outflows and high probability of being magnified (due to strong gravitational lensing by the cluster) because of its spatial location relative to the center of cluster SPT-CL J0640-5113. Another background galaxy that was spectroscopically confirmed in the field of SPT-CL J0313-5334 is an extremely distant Ly$\alpha$ emitter at \textit{z} = 6.15.

\subsection{Redshift uncertainties} \label{unc}

Many of the extracted 1D spectra have significant sky-subtraction residuals. Thus, a differentiation between statistical and systematic errors across the different analysis methods is needed to comprehensively quantify the redshifts.

\begin{figure*}[t!]
\epsscale{1.10}
\centering
\plotone{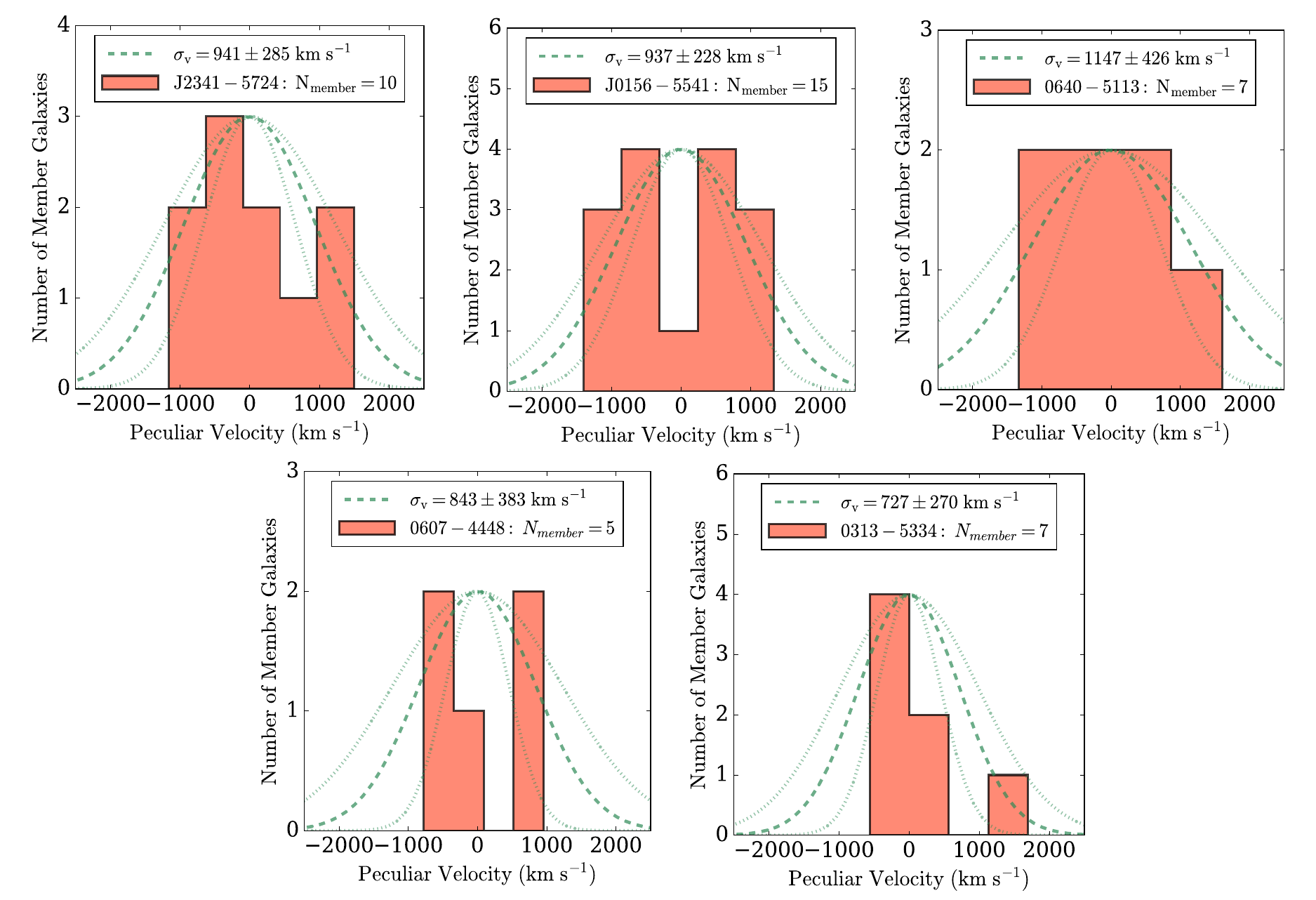}
\caption{Histogram of peculiar velocities (in orange) from the 5 sample clusters (with a total of N=44 galaxies with their respective peculiar velocities). Over-plotted is a Gaussian distribution fit of member galaxy velocities with the standard deviation of $\sigma_{v,G}$ (in dotted green, mean with uncertainties) and a normalization factor corresponding to the maxima of bin counts. See Table \ref{tab:table_total} for details on dispersions and masses of individual galaxy clusters.\label{fig:dispersion}}
\end{figure*}

The median cross-correlation uncertainty reported by RVSAO is $\Delta \textit{z} \sim 10^{-5}-10^{-4}$, whereas the combined median line fit uncertainty (from \textit{customcode}) is $\Delta \textit{z} \sim 10^{-4}-10^{-3}$. The specific value and the ratio of uncertainties from the two methods for an individual spectrum depends on the S/N of the spectrum. Uncertainties in flat-fielding and wavelength calibration for these spectra also contribute to systematic uncertainties ($\Delta \textit{z} \sim10^{-4}$ each). The RVSAO code is known to underpredict uncertanties by at least a factor of 2 even absent any systematic uncertainties \citep{2000AJ....120..511Q, 2016ApJS..227....3B}.

Accounting for systematic uncertainties involved requires a discussion of the RVSAO pipeline. Details of the functioning of RVSAO and physical motivations behind the algorithm are given in \cite{1998PASP..110..934K} and \cite{1979AJ.....84.1511T}, but it is worth revisiting some aspects of the pipeline and choice of parameters that are relevant to the redshift extraction at hand. RVSAO calculates redshifts based on a modified Maximum-Likelihood Estimator, that generates errors based on a cross-correlation peak width obtained from processing an input spectrum. The largest problem in this case is that it is not possible to provide an error-vector to RVSAO i.e. RVSAO assumes every spectral pixel contains a flux value with uniform uncertainty. This limits our ability to interpret RVSAO output uncertainties physically, since our observed spectra have uncertainties that vary significantly with wavelength (as well as with modes in Fourier space), and in the regime of S/N $\sim$1-3.

This is best observed in our analysis if each galaxy spectrum is run through RVSAO over multiple trials, in which most parameters are kept fixed except the following: number of columns in which the data is re-binned in Fourier space, number of times the template spectrum is required to pass through the input galaxy spectrum, wavelength range in which cross-correlation is to be considered, initial redshift guess, and selection cutoffs for Fourier modes to be considered (highest and lowest). The results can be sensitive to these parameters, and the scatter across redshift solutions is expected to reasonably sample the systematic uncertainty. We ran multiple RVSAO trials with all parameters fixed, except cross-correlation wavelength range and initial redshift guess. Each unique wavelength range corresponds to a single trial, used as an input to RVSAO, with different output cross-correlation peaks. Moreover, in trials with relatively small wavelength ranges, care is taken to eliminate wavelength regions of high noise. 

\renewcommand{\arraystretch}{1.0}
\startlongtable
\begin{deluxetable*}{c c c c c c c c}
\centering
\tablecaption{Mean Redshifts, Velocity Dispersions and Mass Comparisons of Sample Galaxy Clusters \label{tab:table_total}}
\tablehead{
\colhead{Cluster Name} & \colhead{Members} & \colhead{$\textit{z}$} & \colhead{$\delta\textit{z}$} & \colhead{$\sigma_{v,G}$\tablenotemark{a}} & \colhead{M$_{200c,SZ}$\tablenotemark{b}} & \colhead{M$_{200c,X-ray}$\tablenotemark{c}} & \colhead{M$_{200c,dyn}$\tablenotemark{d}} \\
\colhead{} & \colhead{(no.)} & & &\colhead{(km s$^{-1}$)} & \colhead{($\times10^{14}$ $h_{70}^{-1}$ M$_\odot$)} & \colhead{($\times10^{14}$ $h_{70}^{-1}$ M$_\odot$)} & \colhead{($\times10^{14}$ $h_{70}^{-1}$ M$_\odot$)}
}
\colnumbers
\startdata 
\textbf{SPT-CL J2341-5724} & 10 & 1.2588 & 0.0021 & $941\pm285$ & ${4.90^{+1.00}_{-1.00}}$ & ${5.40^{+1.20}_{-1.20}}$ & ${5.10^{+5.90}_{-3.30}}$\\
\textbf{SPT-CL J0156-5541} & 15 & 1.2879 & 0.0018 & $936\pm228$ & ${5.90^{+1.20}_{-1.20}}$ & ${6.30^{+1.00}_{-1.00}}$ & ${4.90^{+4.40}_{-2.70}}$\\
\textbf{SPT-CL J0640-5113} & 7 & 1.3162 & 0.0031 &$1147\pm426$& ${5.80^{+1.20}_{-1.20}}$ & ${4.70^{+1.00}_{-1.00}}$ & ${8.80^{+13.20}_{-6.50}}$\\
\textbf{SPT-CL J0607-4448} & 5 & 1.4010\tablenotemark{e} & 0.0028 & $843\pm383$ & ${5.10^{+1.10}_{-1.10}}$ & ${4.30^{+0.90}_{-0.90}}$ & ${3.40^{+6.80}_{-2.80}}$\\
\textbf{SPT-CL J0313-5334} & 7 & 1.4741 & 0.0018 & $727\pm270$ & ${4.90^{+1.10}_{-1.10}}$ & ${3.20^{+2.60}_{-2.50}}$ &${2.20^{+3.20}_{-1.60}}$\\
\enddata
\tablenotetext{a}{Using the robust and resistant gapper estimator, recommended for N$\leq$15 member galaxies, described in \cite{1990AJ....100...32B} and \cite{2014ApJ...792...45R}.}
\tablenotetext{b}{SZ masses reported in B15 and scaled up to M$_{200c}$.}
\tablenotetext{c}{X-ray-temperature based masses reported in \cite{2017arXiv170205094M}.}
\tablenotetext{d}{Using the gapper velocity dispersion, and the M-$\sigma_{v}$ relation (see \citealt{2013ApJ...772...47S} for details).}
\tablenotetext{e}{Cluster redshift determined out of two redshift 'groups', z = 1.40 and z = 1.48. See Section \ref{curiouscase} for more details.}
\end{deluxetable*}

The median scatter in output redshifts observed across multiple trials for the same galaxy spectrum is $\Delta \textit{z} \sim10^{-3}$, which matches the statistical uncertainties obtained from the \textit{customcode} analysis. Keeping the wavelength range and initial redshift guess intact while changing the template spectrum (e.g. SAO's $\it{habtemp90}$) returns a similar range of uncertainties.

As mentioned above, in the presence of these limitations, we quote the most conservative uncertainties for individual redshifts. We start with considering median redshifts from multiple RVSAO cross-correlation trials. For RVSAO, we quote the root mean square (RMS) uncertainties from multiple cross-correlation trials. In the few cases that [OII] emission was observed, we then consider the median of RVSAO cross-correlation and [OII] emission line redshifts, with RMS errors. 

The results from the above analysis are then compared with the \textit{customcode} uncertainties. To be consistent with our approach of reporting the most conservative errors due to presence of unquantifiable sky subtraction systematics, the largest of the three - RVSAO+[OII] uncertainties, \textit{customcode} uncertainties, and the difference in redshift solutions from the two sets of analyses - is taken as the galaxy redshift uncertainty. RVSAO's ability to observe spectral features across different pixel scales (or Fourier modes) in a galaxy spectrum, the agreement in redshifts from three independent analyses, and the confirmation of redshift results by visual inspection of these spectra gives us confidence in our redshift estimates and the characterization of redshift uncertainties. It is also important to note that the scale of redshift uncertainties (or the exact quantitative value for an individual galaxy) in question here does not have a significant effect on the scientific results in this paper, namely the cluster redshifts.

\section{Discussion} \label{discussion}

\subsection{Galaxy Cluster Redshifts (and Velocity Dispersions)} \label{dispersions}
Redshift estimation in this work follows the same procedure as all previous SPT follow-up studies, described in \cite{2014ApJ...792...45R}. It involves using the bi-weight location estimator to calculate the average redshift, $\overline{z}_{cluster}$, assuming a redshift sample drawn from a Gaussian distribution. For the calculation of velocity dispersion, the bi-weight estimator is robust and resistant to outliers and low number statistics. However, in cases of very small samples (N $\leq$ 15), the gapper estimator is preferred, and is used in this work. 
We calculate $\overline{z}_{cluster}$ as best determined using the procedure formulated in \cite{1990AJ....100...32B}. The line-of sight velocity for individual galaxies is computed using the following relationship:

\begin{equation}
v_i = c\frac{(z_i - \overline{z}_{cluster})}{(1 + \overline{z}_{cluster})}
\end{equation}

\begin{figure*}[t!]
\epsscale{1.2}
\centering
\plotone{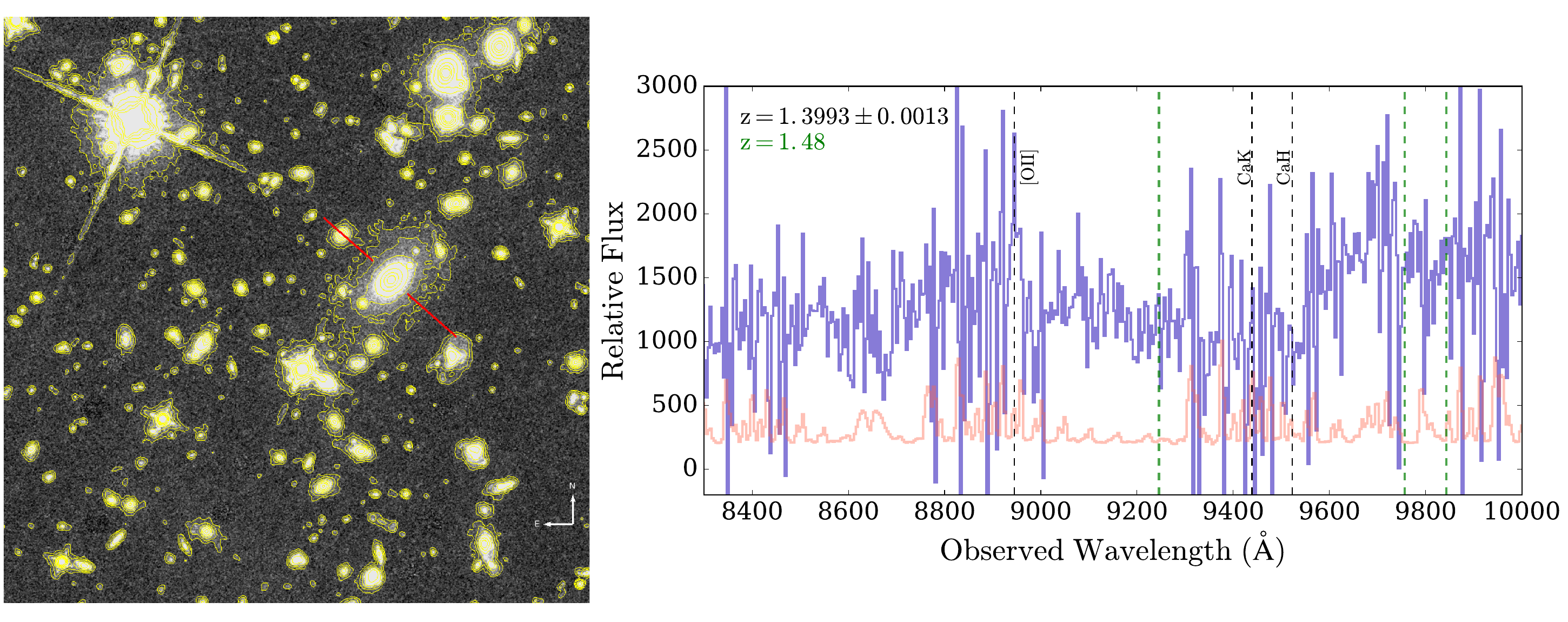}
\caption{{\sl Left}: A 500 kpc cutout of the HST+WFC3 F140W image (Strazzullo et al. in prep) at the cluster redshift \textit{z} = 1.40 for SPT-CL J0607-4448 centered on the SPT-SZ position. Contours are [1.25,2.5,5,10,20,40,60,80,160] times the standard deviation of the sky values, chosen to highlight low-level extended emission seen around galaxies in the image. The galaxy (indicated with red lines) has all the hallmarks of a brightest cluster galaxy (BCG); it is an early type galaxy with an extended stellar halo larger than 100 kpc, and has an appropriate color. {\sl Right}: 1D spectrum for the BCG identified in the left panel. Purple corresponds to observed flux, orange is the 1$\sigma$ error-bars (offset for clarity). Despite absence of a clear diagnostic spectral feature for a redshift, this spectrum favours a \textit{z} = 1.40 solution, based on the vertical green dotted lines corresponding to a redshift Ca II H\&K doublet feature and a feature consistent with [OII] 3727,3729\AA\ emission. The green dotted lines correspond to the same 3 spectral features, but at \textit{z} = 1.48, clearly inconsistent with the spectrum.
\label{fig:spt0607}}
\end{figure*}

\noindent where $\overline{z}_{cluster}$ is the bi-weight location-estimated mean redshift, and the denominator accounts for the difference between the emitter's rest-frame and the cosmological expansion of the universe. The list of velocities $v_i$ is used as an input to the gapper estimator to calculate the line-of-sight velocity dispersion $\sigma_{v}$, once $\overline{z}_{cluster}$ is finalized. This gives us an initial estimate of  $\sigma_{v,G}$. We then account for outliers/interlopers in velocity-space by making a hard $\pm3\sigma$ cut on the distribution of $\sigma_{v,G}$ and ejecting them from the next iteration of calculations until convergence is reached (also see Section \ref{interloper}).
Uncertainties on $\overline{z}_{cluster}$ are calculated using the following expression in \cite{2014ApJ...792...45R} for estimating standard deviation (once again, assuming the measured redshifts are close to a normal distribution):
\begin{equation}
{\Delta}z = \frac{1}{c} \frac{\sigma_{v}(1 + \overline{z}_{cluster})}{\sqrt[]{N_{members}}}
\end{equation}

\noindent where $\sigma_{v} = \sigma_{v,G}$ is the relevant gapper velocity dispersion, 1+$z$ is needed because $\sigma_{v,G}$ is defined in the rest frame, and $1/c$ converts velocity to redshift. Jackknife and bootstrap estimates of this uncertainty also converge to this expression \citep{2014ApJ...792...45R}. Confidence intervals on velocity dispersions are estimated to be:
\begin{equation}
{\Delta}\sigma_{v} = \frac{\pm{C}}{\sigma_{v}\sqrt[]{N_{members} - 1}}
\end{equation}

\noindent This expression accurately captures the confidence interval on the total measurement. For the gapper statistic, C = 0.91.

The final redshifts and velocity dispersions are tabulated in Table \ref{tab:table_total}. Figure \ref{fig:dispersion} shows the distribution of individual cluster member velocities (with an over-plotted Gaussian distribution of mean 0, standard deviation $\sigma_{v,G}$ and an amplitude corresponding to the maxima of the histogram bin counts), where the distribution of member galaxy velocities and the estimated values of the cluster velocity dispersions are consistent with each other.

\begin{figure*}[t!]
\epsscale{1.2}
\plotone{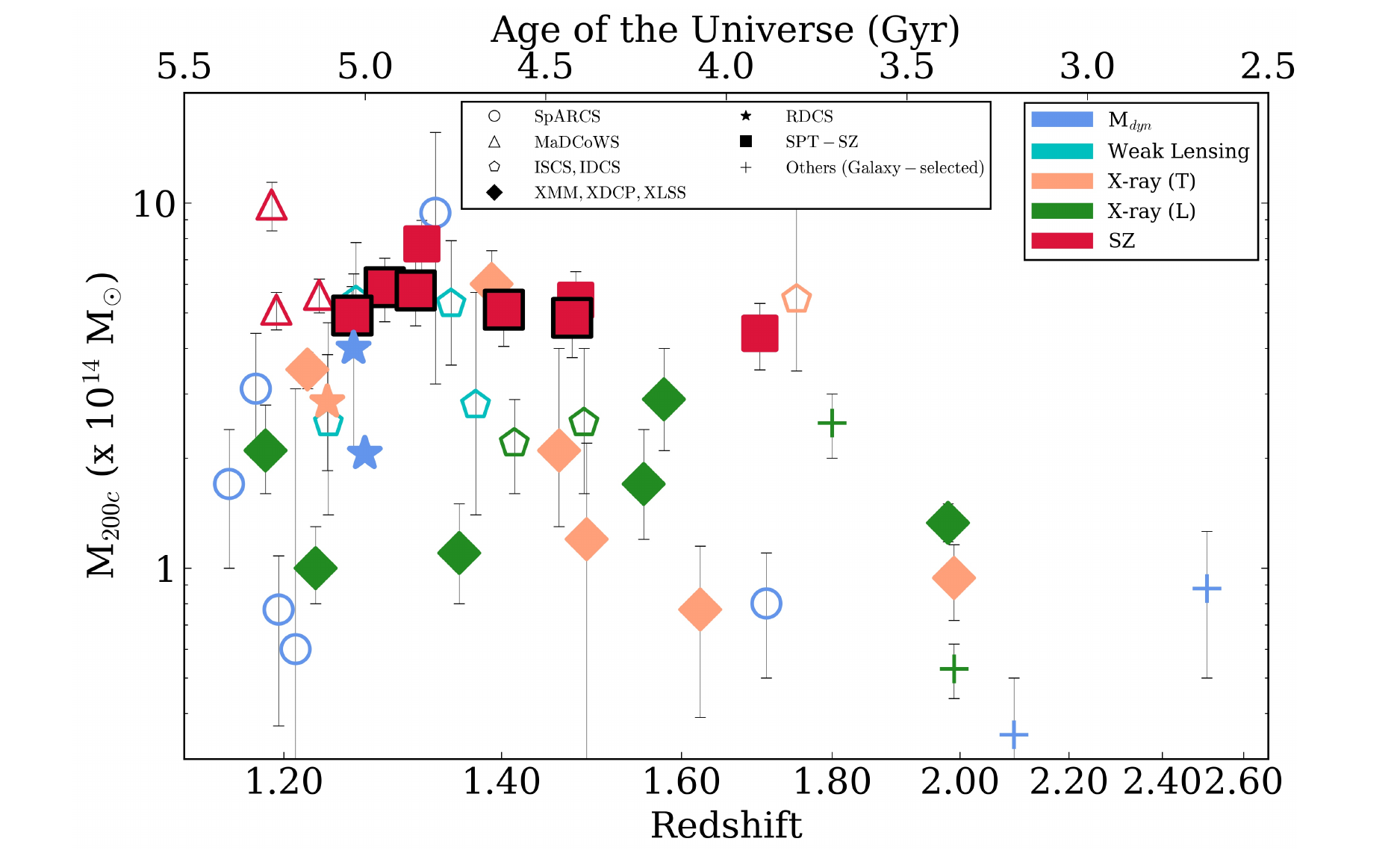}
\caption{Mass vs. redshift (or Age, in Gyr) distribution of all spectroscopically confirmed galaxy clusters with reported masses at \textit{z} > 1.15, including clusters identified in the 2500 deg$^{2}$ SPT-SZ galaxy clusters (\cite{2015ApJS..216...27B}). Red filled squares correspond to the SPT High-z Cluster sample; five clusters with black outline correspond to those analyzed in this work. Also plotted are clusters from major surveys like SpARCS, MaDCoWS, and XMM (marked with different shapes) with their respective cluster mass measurements M$_{200c}$ (marked with different colors). Galaxy luminosity/color-selected clusters are represented by hollow shapes, while ICM-selected clusters are marked with filled shapes in this figure. In cases where M$_{500c}$ is reported, M$_{200c}$ is calculated with the assumptions of an NFW profile and a concentration c$_{500c}$ = 5. 
\label{fig:fig1}}
\end{figure*}

\subsection{The curious case of SPT-CL J0607-4448} \label{curiouscase}

SPT-CL J0607-4448 (\textit{z}$_{phot} = 1.43\pm0.09$, M$_{500c,SZ}\sim 3.14\pm0.64\times10^{14} h^{-1} M_{\odot}$) was targeted for LDSS3 spectroscopy with 20 slits on a multi-object mask, with 10 delivering reliable redshifts (including two field galaxies not associated with the cluster). Eight of the resulting galaxy redshifts were found to be grouped around two redshifts - \textit{z} $\sim 1.40$ (1.4087, 1.4077, 1.3973 and 1.3948), and \textit{z} $\sim 1.48$ (1.4933, 1.4716, 1.4787, and 1.4965). These two redshift groups are separated enough along the Hubble flow that they are certainly distinct objects. However, it is unclear which object dominates the SZ signal that led to the detection of SPT-CL J0607-4448 in the SPT-SZ survey. Based on the properties of four galaxies measured in each redshift group, the velocity dispersions and member galaxy velocity distributions do not clearly favor any one candidate ($\sigma_{v,G}$ for \textit{z}$\sim$1.40 and \textit{z}$\sim$1.48 are 843$\pm$383 and 1587$\pm$834 km s$^{-1}$, respectively). While the \textit{z}$\sim$1.40 dynamics (namely, the numerical value of the velocity dispersion) are relatively more consistent with expectations and the other clusters measured in this paper, the associated large uncertainties need to be noted. The spatial distribution of spectroscopically confirmed galaxies in each group does not indicate a preference for one of the redshifts. However, the detailed photometric analysis of the stellar bump and red sequence colors for SPT-CL J0607-4448 in Strazzullo et al. (in prep) favors the lower redshift solution.

The BCG spectrum (Figure \ref{fig:spt0607}) does not possess a spectral feature (emission or absorption) that produced a clear spectroscopic redshift, due to the presence of particularly strong sky subtraction residuals. Both \textit{customcode} and RVSAO cross-correlation fail to converge to a reliable redshift, but considered against the two choices (\textit{z} = 1.40 or 1.48) the spectrum favors a \textit{z} = 1.40 solution. The black vertical lines correspond to the Ca II H\&K doublet feature redshift to \textit{z} = 1.40, that are in close proximity to a potential 4000\AA\ feature at $\sim$ 9600\AA\ (as opposed to the break presenting itself at $\sim$ 9880\AA\ in the case of a \textit{z} = 1.48 solution). Moreover, there is a potential emission feature at 8944\AA\ that, in isolation, is not compelling, but can be interpreted as [OII] emission at z$\sim$1.3993.  This indicates that the cluster redshift for SPT-CL J0607-4448 is \textit{z} = 1.4010. The galaxies in the \textit{z} $\sim 1.48$ structure (which may or may not be a virialized group or cluster) are noted in Table \ref{tab:table_nonmembers}.


\subsection{The SPT High-z Cluster sample in the context of other clusters in the literature}

This sample contains five high-mass, high redshift SPT-SZ detected galaxy clusters that have been determined photometrically to be above \textit{z} > 1.25. From literature, we find $\sim$ fifty confirmed galaxy clusters at \textit{z} > 1.15, which implies that spectroscopic confirmation of clusters in our sample increases the number of clusters in this regime by 10\%. The SPT high-redshift cluster sample also lies at significantly higher masses than most spectroscopically confirmed clusters at such redshifts; in the high-mass/high-redshift space bounded by the lowest mass and lowest redshift SPT-SZ clusters in this sample, these five spectroscopic confirmations double the total number of confirmed clusters from all previous work.

Figure \ref{fig:fig1} depicts the distribution of M$_{200c}$ as a function of redshift (or elapsed time) of all spectroscopically confirmed galaxy clusters at \textit{z} $> 1.15$ for which masses were reported in literature. The census includes infrared-selected clusters from SpARCS (\citealt{2016A&A...592A.161N},\citealt{2016ApJ...816...48N}), MaDCoWS \citep{2015ApJ...806...26B,2015ApJ...812L..40G} and ISCS \citep{2011ApJ...737...59J,2011ApJ...732...33B,2016ApJ...817..122B} surveys, X-ray selected clusters from XMM \citep{2010ApJ...718...23S}, XDCP \citep{2011NJPh...13l5014F}, XLSS \citep{2015ApJ...811...28T}, and RDCS \citep{1998ApJ...492L..21R} surveys, and SZ-selected clusters from the SPT-SZ survey. This census covers a mass range of $M_{200c}$ $\approx$ 0.3-10 $\times 10^{14}$ M$_\odot$. The colors show the method used to estimate cluster mass: X-ray temperature, X-ray luminosity, SZ, and weak lensing.

The three clusters in red squares without black outlines are SPT-SZ clusters spectroscopically confirmed elsewhere at redshifts greater than 1.2 -- SPT-CL J2040-4451 (\textit{z} = 1.48, \citealt{2014ApJ...794...12B}), SPT-CL J0205-5829 (\textit{z} = 1.32, \citealt{2013ApJ...763...93S}) and SPT-CL J0459-4947 (X-ray spectroscopy-based redshift \textit{z} = 1.70$\pm$0.02, Mantz et al. in prep). The red squares with black outlines represent the 5 SPT high-redshift clusters analyzed in this paper.

It is crucial to note that most of the redshifts confirmed in this work are derived from absorption features despite observational difficulties, while higher redshift clusters are typically confirmed by virtue of strong emission observed e.g. clusters XLSSC 122 ($z=2.0$, \citealt{2017arXiv170308221M}), CL J1001+0220 ($z=2.5$, \citealt{2016ApJ...828...56W}), and the COSMOS-ZFOURGE overdensity ($z=2.1$, \citealt{2014ApJ...795L..20Y}).

\subsection{Validation of Redshift Results}

As previously discussed, obtaining redshifts for primarily passively evolving galaxies at well beyond \textit{z} = 1 is difficult,
and the spectra discussed here are further compromised by systematic sky-subtraction issues. We thus consider in the subsections that follow several analyses of these data beyond cluster redshift estimation, primarily to demonstrate that  the redshifts derived above are consistent with expectations for high redshift clusters.

\subsubsection{Consistency of Dynamical Masses with SZE and X-ray Masses}

We estimate the dynamical masses of these five galaxy clusters using the dispersion-mass scaling relation from \cite{2013ApJ...772...47S}:
\begin{equation}
M_{200c,dyn} = \left(\frac{\sigma_{DM}}{A \times h_{70}(z)^{C}}\right)^{B} 10^{15}M_{\odot}
\end{equation}
where A=939, B=2.91, C=0.33, M$_{200c,dyn}$ is the dynamical mass within R$_{200c}$, defined as the radius within which the mean density $\rho$ is 200 times the critical density $\rho_{c}$ of the universe. $\sigma_{DM}$ is the dispersion computed from galaxy clusters in dark matter simulations, where subhalos correspond to galaxies, while $h_{70}(\textit{z})$ is the redshift-dependent Hubble constant. 

It is assumed here that the average velocity dispersion of galaxies in our clusters can be substituted in the above expression i.e. $\sigma_{DM} \sim \sigma_{v,G}$ , to give a crude estimate of the cluster dynamical masses, which is sufficient given the significant uncertainties associated with velocity dispersion from small numbers of members, and the uncertainty floor imposed by projection and orientation effects in individual clusters \citep{2010MNRAS.408.1818W}.

\begin{figure}[b]
\epsscale{1.2}
\plotone{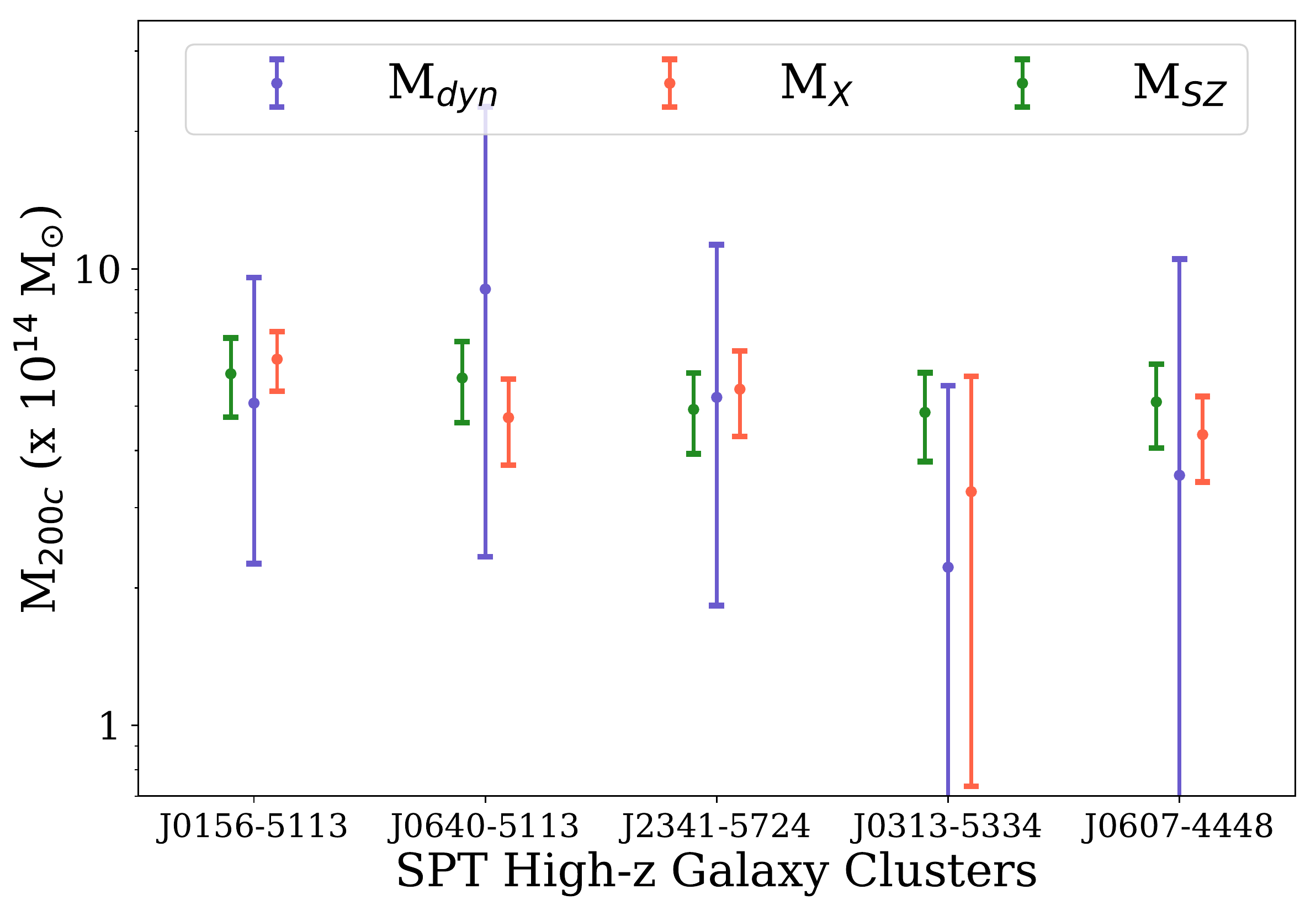}
\caption{M$_{200c}$ comparisons for the SPT High-z cluster sample - dynamical (purple), X-ray (orange) and SZ (green). M$_{200c,dyn}$ calculated in this paper are not inconsistent with other published masses for these galaxy clusters (see Table \ref{tab:table_total} for details).
\label{fig:masses}}
\end{figure}

\begin{figure*}[t!]
\epsscale{1.0}
\centering
\plotone{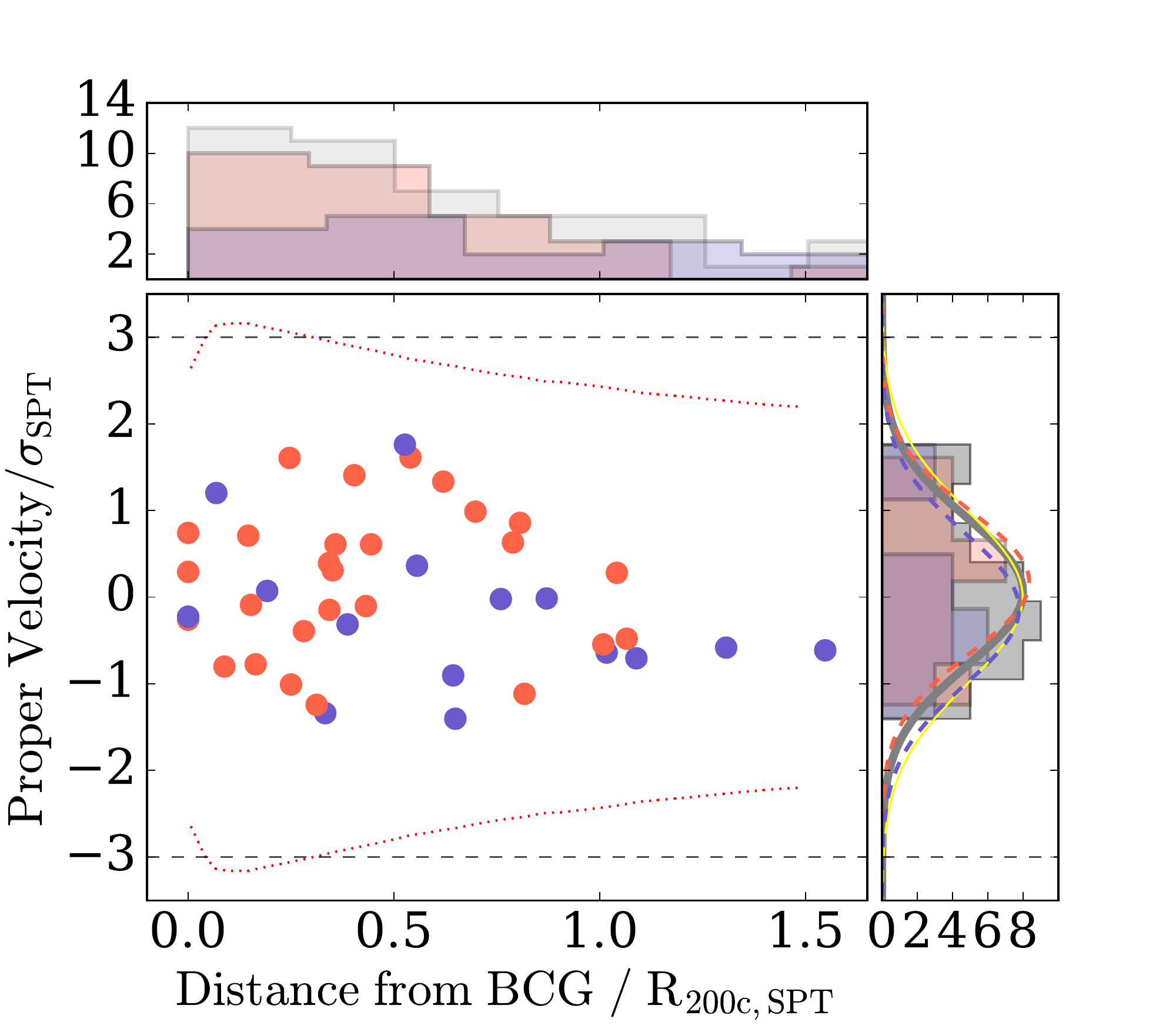}
\caption{Normalized proper velocities vs. normalized distance of member galaxies from SZ center, for the 5 sample clusters (with a total of N = 44 galaxies with their respective peculiar velocities) stacked as a composite cluster. Orange dots represent passive galaxies, and purple dots represent galaxies that exhibit [OII] emission (designated as non-passive galaxies). The red dotted (black dashed) contours represent the radially-dependent $\pm2.7\sigma$(R) (hard $\pm3\sigma$) threshold for interloper rejection. Both velocity and radius histograms show [OII] emitting (purple), passive (orange) and total (grey) population distribution. Over-plotted on the velocity histograms are Gaussian curves corresponding to mean and standard deviation of velocity distributions of the [OII] emitting (dotted purple), passive (dotted orange) and all (solid grey) galaxies. The yellow curve is Gaussian, with a mean of zero and standard deviation of one. The amplitudes for the curves are arbitrary, for pictorial representation. See Table \ref{tab:table_total} for details on dispersions and masses of individual galaxy clusters.
\label{fig:dispersion_stacked_passive}}
\end{figure*}

Additionally, there are potential systematic uncertainties to be considered when comparing the dynamical mass to other estimators, in the conversion from M$_{500c,SZ}$ to M$_{200c,SZ}$ (B15 reports M$_{500c}$); the scale factor is $\sim$ 1.65 in this redshift regime, assuming an NFW profile and a mass-concentration scaling relation from \cite{2008MNRAS.390L..64D}.

Table \ref{tab:table_total} reports the velocity dispersion, the implied dynamical masses, the SZ-derived masses (B15), and the X-ray-temperature derived masses \citep{2017arXiv170205094M} for all five clusters. All masses are reported in M$_{200c}$, scaled where necessary, noting the caveat above. The dynamical mass to SZ mass ratio for this sample (calculated by fitting a line to the M$_{200c,dyn}$--M$_{200c,SZ}$ plane) is $0.73\pm0.36$, and the dynamical mass to X-ray mass ratio is $0.87\pm0.42$. The uncertainty in these ratios is dominated by the high uncertainties in the dynamical masses.

Figure \ref{fig:masses} shows the masses with uncertainties for all five clusters; the dynamical masses are uncertain, but there is no evidence of deviations from expectation that would suggest any systematic issue with the derived galaxy (and in turn, galaxy cluster) redshifts.

\subsubsection{Velocity-Radius Diagrams for a Stacked Cluster} \label{interloper}

As mentioned previously, when calculating the velocity dispersion for a single cluster, we account for outliers/interlopers in velocity-space by making hard $\pm3\sigma$ cuts on the distribution of $\sigma_{v,G}$ and ejecting them from the next iteration of calculations until convergence occurs.

\begin{figure*}[t!]
\epsscale{1.15}
\plotone{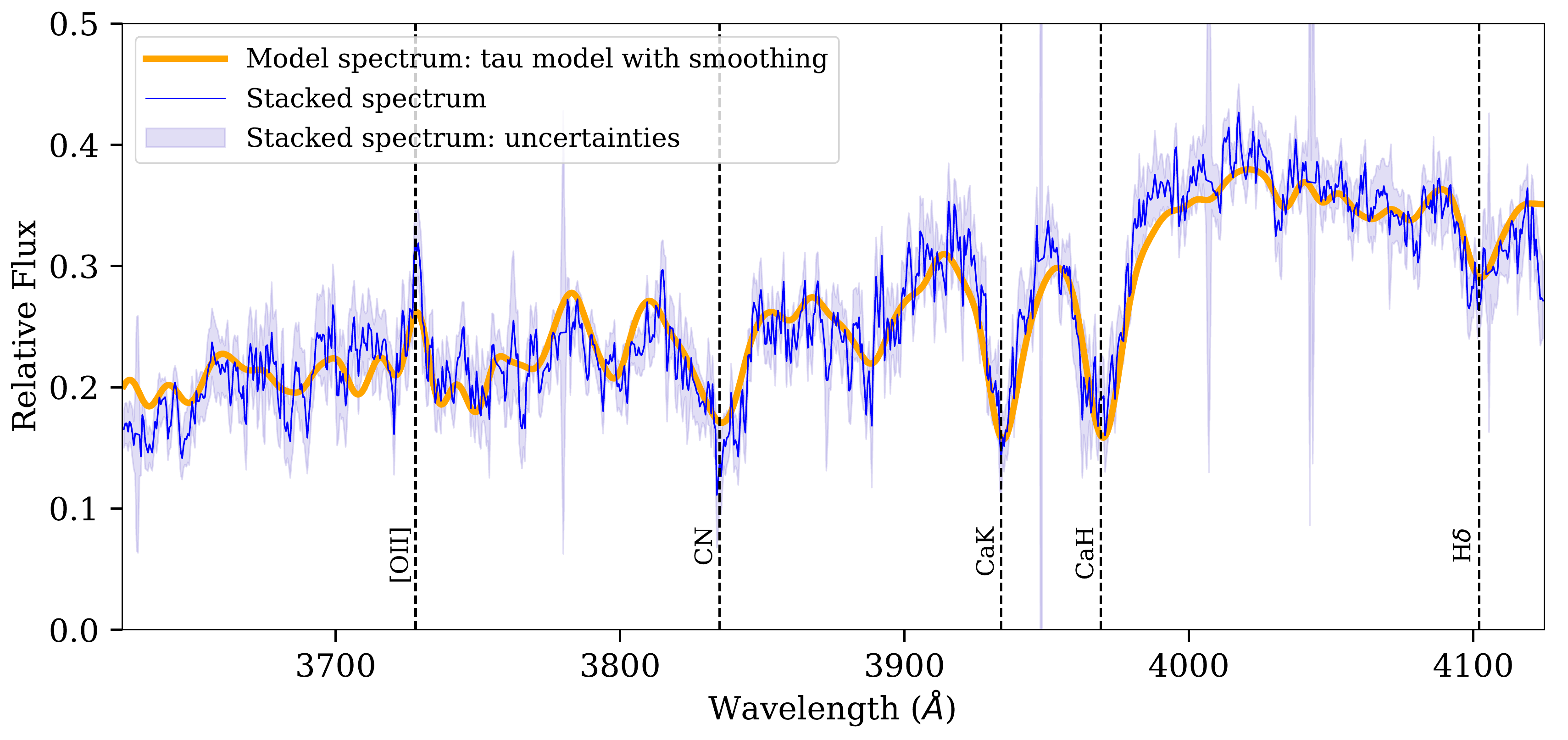}
\caption{Stacked spectrum analysis for the 28 passive galaxies across 5 clusters reported in this paper. The light blue band corresponds to 68\% confidence interval based on a linear combination of statistical and systematic uncertainties in the stack. Orange spectrum corresponds to an exponential tau model of star formation that is 1.7 Gyr old, at a metallicity log (Z/Z$\odot$) = +0.33  (see Section \ref{stack_prospector} for a more detailed description). Dotted lines correspond to rest frame [OII], CN, Ca II H\&K, and H$\delta$ features. 
\label{fig:spectra_stacked}}
\end{figure*}

To examine this phase space further, we create a stacked cluster from the composite distribution of all 44 member galaxy velocities and galaxy distances from the SZ centers. The SZ mass is used to normalize velocities by an equivalent dispersion $\sigma_{200c,SZ}$, calculated using the dispersion-mass scaling relation \citep{2013ApJ...772...47S} from M$_{200c,SZ}$ (scaled up from the SPT mass M$_{500c,SZ}$) akin to the previous section. The projected radial distances of individual member galaxies are also normalized by R$_{200c,SPT}$. The resulting phase-space diagram is shown in Figure \ref{fig:dispersion_stacked_passive}, along with the peculiar velocity distribution in the `stacked cluster'. The horizontal dotted lines correspond to $\pm3\sigma$ threshold, while the orange dotted curve is the radially dependent $\pm2.7\sigma$(R) threshold from an NFW profile, for optimal interloper rejection \citep{2010A&A...520A..30M}. From Figure \ref{fig:dispersion_stacked_passive}, we conclude that: a) the simple 3-sigma outlier rejection used in Section \ref{dispersions} is sufficient, and b) the radial profile of velocities in the stacked cluster look as expected (i.e. small at the center, rising to a maximum, and decreasing at large radii), suggesting that cluster members have been robustly measured and identified.

In addition, we run a Kolmogorov-Smirnov (K-S) test on the total galaxy population's velocity distribution (gray horizontal histogram, Figure \ref{fig:dispersion_stacked_passive}). The K-S statistic is 0.08 i.e. it does not reject the hypothesis that normalized galaxy velocities in our cluster sample are drawn from a Gaussian distribution. This is consistent with expectations (see \citealt{2014ApJ...792...45R,2017ApJ...837...88B}). 

We also analyze the distribution of cluster member galaxies in velocity-radius phase space by distinguishing passive (orange) from [OII]-emitting (purple) galaxies. Nominally passive galaxies describe a more centrally condensed distribution by comparison to the more extended distribution of galaxies exhibiting [OII] emission. This is likely a real trend and unlikely to be a simple selection effect - placing slits on bright red apparent cluster galaxies at larger radii is easier than in the cluster center due to less crowding, and there are potential red cluster members at all radii in the imaging data. Moreover, it is seen that the ratio of passive galaxy to [OII]-emitting galaxy velocity dispersion is 0.95$\pm$0.26, in good agreement with trends observed by \cite{2017ApJ...837...88B} for low- and medium-redshift SPT-discovered galaxy clusters. This projected radius and velocity segregation between passive and emission-line galaxies is thought to indicate differences in formation timescales and accretion histories into the cluster environment. That the entire galaxy population of the stacked cluster when dissected in this manner is again consistent with expectations from lower-redshift clusters also indicates that cluster member redshifts have been well measured. 

\subsection{Stacked spectral analysis of Passive Galaxies}\label{stack_prospector}

We construct a composite spectrum of 28 passive member galaxies across 5 clusters, i.e. all galaxies for which an [OII] emission feature was not detected. To stack, we shift each spectrum to the rest frame (based on their final reported redshift), and map it to the wavelength range 3645-4125\AA, with a flux normalization using the nominal throughput curve for the instrument LDSS3-C in this configuration. This is followed by a weighted sum stacking of the 28 spectra, where each flux value corresponding to a wavelength is weighted by the error vector for each galaxy spectrum. We further excluded a portion of each spectrum from the stack; the excluded data are any pixels with nominal uncertainties greater than 2x the mean uncertainty of the ten pixels with the lowest uncertainty in each input spectrum. This typically excludes about 30\% of the input pixels, which correspond in each instance to the majority of the pixels that have large sky subtraction residuals. A systematic uncertainty is calculated by varying the exclusion percentage upward and downward by 10\% (i.e., typically from 20\%-40\% of the pixels are excluded) in steps of 1\%, and computing a stacked spectrum at each cut. The variance at each pixel across the resulting 21 different stacks is taken as an estimate of systematic uncertainty. The statistical uncertainty is calculated by bootstrapping the spectra input to the stacking process. The final reported uncertainty is the sum in quadrature of the systematic and statistical uncertainties, which typically are of comparable magnitude. The stacked spectrum (blue, with the 68\% confidence interval in light blue) is shown in Figure \ref{fig:spectra_stacked}. 

Notably, in the stacked spectra, we detect a composite [OII] emission feature, not previously detected in individual spectra. Additionally the spectrum clearly shows a pronounced broad CN feature in the range of 3820-3850\AA, as well as H$\delta$ absorption at 4102\AA. We also perform stellar population synthesis modeling with our stacked spectrum using the MCMC code Prospector \citep{2017ApJ...837..170L,ben_johnson_2017_1116491,2013PASP..125..306F,2010ApJ...712..833C} to demonstrate that the aggregate spectrum is reasonable and as expected for cluster member galaxies at this epoch. In Figure \ref{fig:spectra_stacked}, we overplot a best-fit spectrum using a simple tau ($\tau$) model (e-folding time = 300 Myr, in orange) for a 1.7 Gyr old stellar population, at a metallicity log (Z/Z$\odot$) = 0.33 with a velocity broadening over scales of 275 km s$^{-1}$. Dotted lines correspond to rest frame [OII], CN, Ca II H\&K, and H$\delta$ features. The clear emergence of [OII], H$\delta$ and CN features - which were not used to establish redshifts for any of these galaxies - and the overall good correspondence between the stacked and the quite reasonable model spectrum, is yet one more validation of the redshifts of the individual galaxies that were used in the composite stacking.
A comprehensive analysis of physical properties of stellar populations in the cluster members characterized here shall be presented in a future paper (Khullar et al., in prep).

\section{Summary} \label{summary}

We present spectroscopic follow-up of 5 of the most distant galaxy clusters in the 2500 deg$^{2}$ SPT-SZ survey - part of the SPT High-z Cluster sample. This work describes the observations, the spectroscopic analysis pipeline, and the data products that have been subsequently derived. We analyze this data set via cross-correlation, and manual emission and absorption line fits, to infer robust spectroscopic redshifts for member galaxies. We argue that despite the presence of mostly low S/N spectra dominated by sky background noise (associated with sky subtraction residuals, an artifact of the data quality and the reduction process), useful parameters can be extracted from the dataset. We perform several consistency checks for the reported spectroscopic redshifts - calculations of velocity dispersions and dynamical masses, exploration of the velocity-radius phase space for cluster member galaxies, and a composite stacked spectrum that exhibits features of nominally passive galaxies. The reported set of galaxy cluster redshifts doubles the number of galaxy clusters spectroscopically confirmed at M$_{200c}$ $\geq 4.5\times10^{14} M_{\odot} h^{-1}$ and at \textit{z} > 1.2. 

This work has been an effort to spectroscopically characterize the highest redshift massive galaxy clusters from the SPT-SZ catalog. The distant, massive cluster population presented in this work represents the progenitors of nearby massive clusters; as such it is imperative to study this sample both observationally and in comparison with simulations. Despite limitations in spectral observations (mostly pertaining to quantifying systematics in sky subtraction), as this work reports robust cluster redshifts, future spectroscopy of these distant and faint clusters would be able to employ techniques with optimal sky subtraction (e.g. nod-and-shuffle mode on Magellan/LDSS3 targets a narrower spectral range but is an improved handling of systematics; see \citealt{2001PASP..113..197G}). This spectroscopic confirmation study encourages further follow-up that targets observations of star formation rates and history, tracers of cluster dynamics, and estimation of velocity segregation and biases in these unique systems.

\section{Acknowledgments}
\acknowledgements

GK thanks Chihway Chang, Andrey Kravtsov, Richard Kron and Huan Lin, for their helpful and thoughtful feedback that improved the analysis in this paper. This paper has gone through internal review by the South Pole Telescope collaboration.

This work is supported by the Department of Astronomy and Astrophysics at the University of Chicago, NSF Physics Frontier Center grant PHY-1125897 to the Kavli Institute of Cosmological Physics at the University of Chicago, as well as by the Kavli Foundation, and the Gordon and Betty Moore Foundation grant GBMF 947. The South Pole Telescope is supported by the National Science
Foundation through grant PLR-1248097.

BB is supported by Fermi Research Alliance, LLC under Contract No. DE-AC02-07CH11359 with the U.S. Department of Energy, Office of Science, Office of High Energy Physics. Argonne National Laboratory work was supported under U.S. Department of Energy contract DE-AC02-06CH11357. MB was supported by National Science Foundation through Grant AST-1009012. AS is supported by the ERC-StG `ClustersXCosmo', grant agreement 71676. The data analyzed in this paper was taken on the 6.5m Magellan Telescopes at the Las Campanas Observatory, Chile, supported by the Carnegie Observatories. This work is partly based on observations made with the NASA/ESA Hubble Space Telescope, obtained at the Space Telescope Science Institute, which is operated by the Association of Universities for Research in Astronomy, Inc., under NASA contract NAS 5-26555, associated with SPT follow-up GO program 14252. This work is based in part on observations made with the {\sl Spitzer Space Telescope}, which is operated by the Jet Propulsion Laboratory, California Institute of Technology under a contract with NASA. CR acknowledges support from Australian Research Council's Discovery Projects scheme (DP150103208).

\facilities{SPT, Magellan:Clay (LDSS3, PISCO), Baade (IMACS, FOURSTAR), Chandra, Spitzer, HST} 
\software{COSMOS, Python - Numpy, Scipy, Astropy \citep{2018arXiv180102634T}, Colossus, Matplotlib, Pandas, Python-fsps, Prospector, emcee, IRAF- NOAO,RVSAO}

\bibliographystyle{yahapj}
\bibliography{references}

\end{document}